\documentclass[fleqn,usenatbib]{mnras}
\usepackage[T1]{fontenc}
\usepackage{ae,aecompl}
\usepackage{subfigure}
\usepackage{graphicx}	
\usepackage{amsmath}	
\usepackage{amssymb}	
\usepackage[dvipsnames]{xcolor}
\usepackage{hyperref}
\usepackage{lscape}
\usepackage{multirow}
\usepackage{booktabs, caption, makecell}
\usepackage[para,online,flushleft]{threeparttable}
\usepackage{longtable}
\usepackage{changepage}
\usepackage{ulem}
\usepackage{lipsum}
\usepackage{bm}
\newcommand{\etal}{\textit{et al. }}
\def\CIVdblt{{\rm C~}\kern 0.1em{\sc iv}~$\lambda\lambda 1548, 1550$}
\def\NVdblt{{\rm N~}\kern 0.1em{\sc v}~$\lambda\lambda 1238, 1242$}
\def\OVIdblt{{\rm O~}\kern 0.1em{\sc vi}~$ 1031, 1037$}
\def\SVIdblt{{\rm S~}\kern 0.1em{\sc vi}~$ 933, 944$}
\def\SiIVdblt{{\rm Si~}\kern 0.1em{\sc iv}~$\lambda\lambda1394, 1403$}
\def\MgIIdblt{{\rm Mg~}\kern 0.1em{\sc ii}~$\lambda\lambda2796, 2803$}
\def\NeVIIIdblt{{\rm Ne~}\kern 0.1em{\sc viii}~$\lambda\lambda770, 780$}
\def\NeV{\hbox{{\rm Ne~}\kern 0.1em{\sc v}}}
\def\NeVI{\hbox{{\rm Ne~}\kern 0.1em{\sc vi}}}
\def\NeVIII{\hbox{{\rm Ne~}\kern 0.1em{\sc viii}}}
\def\OII{\hbox{{\rm O~}\kern 0.1em{\sc ii}}}
\def\OIII{\hbox{{\rm O~}\kern 0.1em{\sc iii}}}
\def\OIV{\hbox{{\rm O~}\kern 0.1em{\sc iv}}}
\def\OV{\hbox{{\rm O~}\kern 0.1em{\sc v}}}
\def\OVI{\hbox{{\rm O~}\kern 0.1em{\sc vi}}}
\def\OVII{\hbox{{\rm O~}\kern 0.1em{\sc vii}}}
\def\OVIII{\hbox{{\rm O~}\kern 0.1em{\sc viii}}}
\def\NI{\hbox{{\rm N~}\kern 0.1em{\sc i}}}
\def\NII{\hbox{{\rm N~}\kern 0.1em{\sc ii}}}
\def\NIII{\hbox{{\rm N~}\kern 0.1em{\sc iii}}}
\def\NIV{\hbox{{\rm N~}\kern 0.1em{\sc iv}}}
\def\NV{\hbox{{\rm N~}\kern 0.1em{\sc v}}}
\def\NVII{\hbox{{\rm N~}\kern 0.1em{\sc vii}}}
\def\CII{\hbox{{\rm C~}\kern 0.1em{\sc ii}}}
\def\CIII{\hbox{{\rm C~}\kern 0.1em{\sc iii}}}
\def\SiII{\hbox{{\rm Si~}\kern 0.1em{\sc ii}}}
\def\SiIII{\hbox{{\rm Si~}\kern 0.1em{\sc iii}}}
\def\SIV{\hbox{{\rm S~}\kern 0.1em{\sc iv}}}
\def\SV{\hbox{{\rm S~}\kern 0.1em{\sc v}}}
\def\SVI{\hbox{{\rm S~}\kern 0.1em{\sc vi}}}
\def\SiI{\hbox{{\rm Si~}\kern 0.1em{\sc i}}}
\def\PII{\hbox{{\rm P~}\kern 0.1em{\sc ii}}}
\def\AlII{\hbox{{\rm Al~}\kern 0.1em{\sc ii}}}
\def\AlIII{\hbox{{\rm Al~}\kern 0.1em{\sc iii}}}
\def\CaI{\hbox{{\rm Ca~}\kern 0.1em{\sc i}}}
\def\CaII{\hbox{{\rm Ca~}\kern 0.1em{\sc ii}}}
\def\CrII{\hbox{{\rm Cr~}\kern 0.1em{\sc ii}}}
\def\CII{\hbox{{\rm C~}\kern 0.1em{\sc ii}}}
\def\CIII{\hbox{{\rm C~}\kern 0.1em{\sc iii}}}
\def\CIV{\hbox{{\rm C~}\kern 0.1em{\sc iv}}}
\def\CV{\hbox{{\rm C}\kern 0.1em{\sc v}}}
\def\MgX{\hbox{{\rm Mg}\kern 0.1em{\sc x}}}
\def\MgII{\hbox{{\rm Mg}\kern 0.1em{\sc ii}}}
\def\FeII{\hbox{{\rm C~}\kern 0.1em{\sc ii}}}
\def\FeIII{\hbox{{\rm C~}\kern 0.1em{\sc iii}}}
\def\H{\hbox{{\rm H~}}}
\def\HI{\hbox{{\rm H~}\kern 0.1em{\sc i}}}
\def\HeI{\hbox{{\rm He~}\kern 0.1em{\sc i}}}
\def\HII{\hbox{{\rm H~}\kern 0.1em{\sc ii}}}
\def\Lya{\hbox{{\rm Ly}\kern 0.1em$\alpha$}}
\def\Lyb{\hbox{{\rm Ly}\kern 0.1em$\beta$}}
\def\Lyg{\hbox{{\rm Ly}\kern 0.1em$\gamma$}}
\def\Lyth{\hbox{{\rm Ly}\kern 0.1em$\theta$}}
\def\Lyfive{\hbox{{\rm Ly}\kern 0.1em$5$}}
\def\Lysix{\hbox{{\rm Ly}\kern 0.1em$6$}}
\def\Lyseven{\hbox{{\rm Ly}\kern 0.1em$7$}}
\def\Lyeight{\hbox{{\rm Ly}\kern 0.1em$8$}}
\def\Lynine{\hbox{{\rm Ly}\kern 0.1em$9$}}
\def\Lyten{\hbox{{\rm Ly}\kern 0.1em$10$}}
\def\MnII{\hbox{{\rm Mn~}\kern 0.1em{\sc ii}}}
\def\kms{\hbox{km~s$^{-1}$}}
\def\cmsq{\hbox{cm$^{-2}$}}
\def\cc{\hbox{cm$^{-3}$}}
\newcommand{\angstrom}{\mbox{\normalfont\AA}}

\usepackage{soul}
\usepackage{tabularx,ragged2e}
\usepackage{booktabs,array,times}
\usepackage[export]{adjustbox}
\usepackage{siunitx}
\usepackage{rotating}

\title[Pair Lines of Sight Observations of Multiphase Gas]{Pair Lines of Sight Observations of Multiphase Gas Bearing {\OVI} in a Galaxy Environment 
}

\author[Anshul et al.]{Pratyush Anshul$^{1}$,\thanks{E-mail: anshulpratyush7@gmail.com}
Anand Narayanan,$^{1}$
Sowgat Muzahid,$^{2}{}^{,3}$
Alexander Beckett$^{4}{}^{,5}$ and
\newauthor Simon L. Morris$^{4}{}^{,5}$
\\
$^{1}$Department of Earth and Space Sciences, Indian Institute of Space Science \& Technology, Thiruvananthapuram 695547, Kerala, India\\
$^{2}$IUCAA, Post Bag 04, Ganeshkhind, Pune-411007, India\\
$^{3}$Leibniz-Institut fur Astrophysik Potsdam (AIP), An der Sternwarte 16, D-14482 Potsdam, Germany\\
$^{4}$Department of Physics, Durham University, South Road, Durham DH1 3LE, UK\\
$^{5}$Centre for Extragalactic Astronomy, Durham University, South Road, Durham DH1 3LE, UK
}
\DeclareUnicodeCharacter{2212}{-}
\begin{document}
\label{firstpage}
\pagerange{\pageref{firstpage}--\pageref{lastpage}}
\maketitle

\begin{abstract}

Using $HST$/COS observations of the twin quasar lines of sight Q~$0107-025$A \& Q~$0107-025$B, we report on the physical properties, chemical abundances and transverse sizes of gas in a multiple galaxy environment at $z = 0.399$ across a transverse separation of $520$~kpc. The absorber towards Q~$0107-025$B has $\log N(\HI)/\cmsq \approx 16.8$ (partial Lyman limit) while the absorber towards the other sightline has $N(\HI) \approx 2$ dex lower. The {\OVI} along both sightlines have comparable column densities and broad $b$-values, whereas the low ionization lines are considerably narrower. The low ionization gas is inconsistent with the {\OVI} when modelled assuming photoionization in a single phase. Along both the lines-of-sight, {\OVI} and coinciding broad {\HI} are best explained through collisional ionization in a cooling plasma with solar metallicity. Ionization models infer $1/10$-th solar metallicity for the pLLS and solar metallicity for the lower column density absorber along the other sightline. Within $\pm~250$~{\kms} and $2$~Mpc of projected distance from the sightlines 12 galaxies are identified, of which 3 are within $300$~kpc. One of them is a dwarf galaxy while the other two are intermediate mass systems at impact parameters of $\rho \sim (1-4)~R_{vir}$. The O VI along both lines-of-sight could be either tracing narrow transition temperature zones at the interface of low ionization gas and the hot halo of nearest galaxy, or a more spread-out warm gas bound to the circumgalactic halo/intragroup medium. This latter scenario leads to a warm gas mass limit of $M \gtrsim 4.5 \times 10^{9}$~M$_\odot$.

\end{abstract}



\section{Introduction}

Our understanding of the distribution and physical properties of diffuse baryons in the universe has primarily come from quasar absorption line studies and deep galaxy redshift surveys. The technique remains one of the most sensitive ways to gain insights on gas in a wide variety of environments, from the dense interstellar regions in galaxies to the diffuse filamentary and sheet like structures that form the intergalactic cosmic web \citep{1996Natur.380..603B}. However, such studies are routinely based on single line of sight observations, with the physical description of the absorber drawn solely from the one dimensional information along the line of sight. This can be circumvented partially by collecting a statistically large sample of similar absorption systems along several random lines of sight \citep[e.g.,][]{2012ApJ...750...67R,2013ApJ...777...59T}. But even there, the spatial information on the different gas phases, their volume filling fraction within the virial radius, and the baryonic mass they encompass are all contingent on some assumed geometry based on theoretical considerations. 

\begin{table*}
    \centering
    \caption{Details of the triplet quasars taken from \citet{2010MNRAS.402.1273C} and \citet{2010A&A...520A.113B}.}
    \label{tab:quasar details}
    \begin{tabular}{clccccr}
        \hline
        Sightline ID & QSO & RA (J2000) & DEC (J2000) & $z_{em}$ & R (mag) & V (mag) \\
        \hline
        A & LBQS0107-025A (Q0107-025A) & 01:10:13.14 & -02:19:52.9 & 0.960 & 18.1 & 17.8\\
        B & LBQS0107-025B (Q0107-025B) & 01:10:16.25 & -02:18:51.0 & 0.956 & 17.4 & 17.3\\
        C & LBQS0107-0232 (Q0107-0232) & 01:10:14.43 & -02:16:57.6 & 0.726 & 18.4 & \\
        \hline
    \end{tabular}
\end{table*}

An approach to overcome this limitation is to observe multiple closely separated lines of sight to background quasars through the same foreground absorber \citep[e.g.,][]{1994ApJ...437L..83B,1995Natur.373..223D,1997ApJ...491...45D,1999ApJ...513..598L,2000A&A...357...37L,2001ApJ...562...76R,2006AJ....132.2046P,2010MNRAS.402.1273C,2019ApJ...885...61R}. Spatially resolved spectroscopy of gravitationally lensed quasar images or binary quasars are ideal for such close-by transverse lines of sight studies. By resolving the transverse sizes, density, temperature and metallicities at kiloparsec and smaller scales, these observations have resulted in direct insights into the geometrical distribution of gas of various ionization states in a wide variety of absorbing environments \citep{1998ApJ...502...16C,1999ApJ...515..500R,2005ApJ...626..767L,2013AJ....145...48M,2014ApJ...784....5M,2020ApJ...900....9L}.

For example, comparisons of {\HI} absorption profiles across pairs of lines of sight, and also  cross-correlation measures of transmitted fluxes between multiple closely separated lines of sight have shown {\Lya} absorbers of the IGM to be coherent gas structures with sizes ranging from $200$~kpc to $1$~Mpc \citep[e.g.,][]{1998A&A...334L..45P,2002A&A...391....1A,2004ApJ...613...61B}. Such estimates have served as key constraints in developing a three dimensional structure of the cosmic web \citep{2010MNRAS.407.1290C} and in its comparison with theoretical expectations. Similarly, the structural properties of low and high ionization gas phases in metal absorbers have also been investigated through transverse lines of sight studies. Ionization studies of {\MgII} absorption systems had long suggested that they constitute a two-phase medium with the {\MgII} tracing dense ($n_{\H} > 10^{-3}$~{\cc}) clouds with sizes of a few hundred parsec (assuming spherical geometry), and the {\CIV} that is often coincident in redshift with the {\MgII}, tracing a more diffuse ($n_{\H} \sim 10^{-4}$~{\cc}) higher ionization envelope surrounding the low ionization {\MgII} gas \citep{1986A&A...169....1B,1990ApJS...72....1S,1999ApJ...519L..43C,2000ApJS..130...91C}. Through spectroscopic observations of gravitationally lensed quasar images, \citet{1999ApJ...515..500R,2001ApJ...562...76R,2002ApJ...576...45R} and \citet{2002ApJ...569..676K} were able to corroborate these claims by showing that the {\MgII} absorption remains coherent over transverse scales of $200 - 400$~pc or smaller whereas the associated {\CIV} can extend over scales of $\gtrsim 10$~kpc, but with variations at sub-kiloparsec scales \citep{2018ApJ...859..146R}. The compact nature of low ionization gas thus established also explains why the redshift number densities of {\MgII} are lower compared to {\CIV} systems in spectral line surveys of comparable sensitivity. 

Similar studies involving {\OVI} systems have helped to constrain the distribution of higher ionization warm/hot gas in the CGM of galaxies. The {\OVI} absorption is found to be uniform over larger transverse scales than {\CIV} suggesting that the highly ionized halos of galaxies are more extended than the photoionized {\CIV} reveals \citep{1999ApJ...515..500R,2007A&A...469...61L}. By following {\OVI} absorption along and across closely separated lines of sight, \citet{2014ApJ...784....5M} was able to estimate a size of $R \sim 330$~kpc for the ionized CGM surrounding a $1.2L^*$ galaxy, with a baryonic mass of $M \sim 1.2 \times 10^{11}$~M$_{\odot}$. In the absence of information on the transverse sizes, it is difficult to arrive at such estimations without assuming a morphology for the absorbing medium. More recently, using 43 background QSO lines of sight, \citet{2020ApJ...900....9L} were able to derive detailed information on the distribution of multiphase gas in the extended halo of M31 from close to the galactic disk ($R = 25$~kpc) to nearly twice its virial radius ($R = 570$~kpc). The absorption along multiple sightlines showed a nearly $100$\% covering fraction for {\OVI} in comparison to all other ionic species, out to $\sim 1.9R_{vir}$, revealing the wide extend of the diffuse warm-hot CGM around M31 \citep[see also][]{2015ApJ...804...79L}. 
\begin{figure*}
    \begin{center}
    \includegraphics[width=0.8\textwidth]{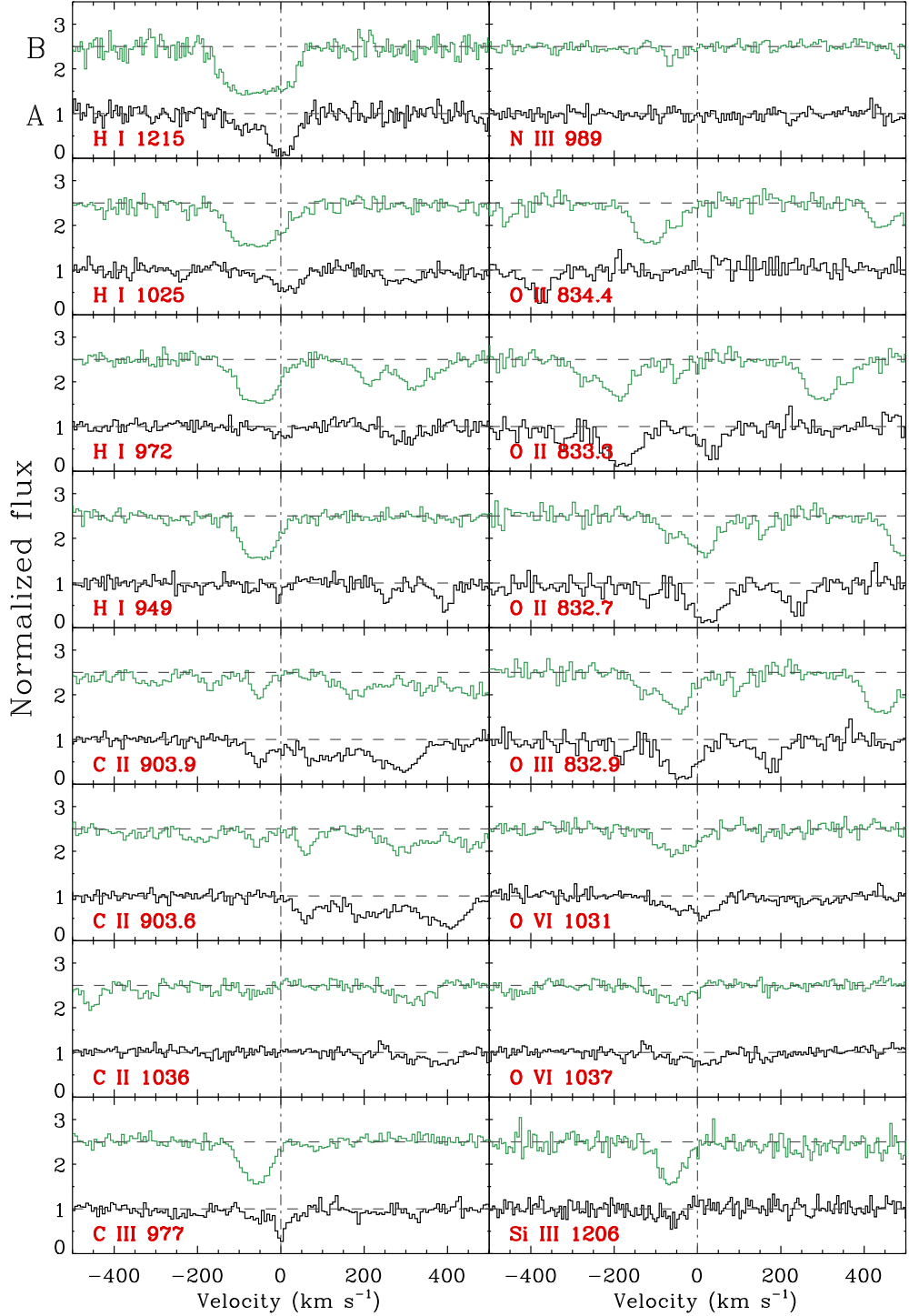}
    \caption{Velocity plots of the key transitions towards sightline A (black) and sightline B (green). The Y-axis is continuum normalized flux, and the X-axis is velocity along the line-of-sight in the rest frame of the absorber. Zero velocity corresponds to z $= 0.39949$. The sightline B spectrum is arbitrarily shifted along the vertical axis for clarity. Observed wavelengths of the different transitions are indicated in the different panels. Voigt profile fits to the detected lines along the sightline A is shown in Fig.~\ref{fig:A_detections} and along the sighline B is shown in Fig.~\ref{fig:B_detections_BLA} and Fig.~\ref{fig:B_detections_metals}.}
    \label{fig:combined_spectra}
\end{center}    
\end{figure*}

In this work, we present analysis on intervening {\OVI} absorption detected along the lines of sight to the quasars Q$0107-025$A, and Q$0107-025$B, separated by $1.29^{\prime}$ in the plane of the sky. The $HST$/COS spectra of these twin lines of sight show O VI absorbers at $z = 0.227$, and $z = 0.399$. The former absorber was analyzed by \citet{2014ApJ...784....5M} who found identical ionization and chemical properties for the {\OVI} bearing high ionization gas, indicating that the lines of sight are probing a uniform large-scale structure, corresponding to the highly ionized CGM of a luminous galaxy. Here we present a similar analysis for the $z = 0.399$ {\OVI} absorber with information on the distribution of galaxies in the extended environment surrounding the absorber. The paper is organized as follows: the details on the COS spectra, and galaxy MOS data are described in Section 2 and Section 3, respectively. The spectral line analysis and ionization models are dealt with in Section 4. The association of the absorbers with galaxies, and the summary discussion are given in Section 5 and 6, respectively. Throughout the paper, we adopt the cosmology with H$_{0}$ = $69.6~km~s^{-1}~Mpc^{-1}$, $\Omega_m = 0.286$ and $\Omega_{\Lambda} = 0.714$ from \citet{2014ApJ...794..135B}. All the logarithmic values mentioned are in base-10. The distances mentioned in this paper are physical unless stated otherwise.

\section{COS observations of the Quasar Pair}

The quasars Q$0107-025$A ($z_{em} = 0.960$) \& Q$0107-025$B ($z_{em} = 0.956$) (hereafter sightlines A \& B) were discovered as triplet of quasars by \citet{1983LIACo..24..355S,1986A&A...161..209S}. Their coordinates, spectroscopic emission redshifts, and apparent brightness in the V-band are given in Table \ref{tab:quasar details}. The sight lines have a projected angular separation of $1.29^{\prime}$ in the plane of the sky. The $HST$/COS G130M and G160M grating observations of these targets were carried out under the Program ID: 11585 (PI: Neil Crighton), with a total integration time of $\approx 11.6$~ks on each quasar. The final co-added spectra of A \& B sight lines were downloaded from the $HST$ Spectroscopic Legacy Archive \citep{2017cos..rept....4P}. The spectra span the far-UV wavelength range from $1150 - 1775$~{\AA} with a spectral resolution of $\sim 17 - 20$~{\kms} per resolution element of 0.06~{\AA}. Continuum normalization was done by fitting lower order polynomials to the line free regions. The two quasars are part of a triplet system that includes the quasar Q$0107-0232$ at $z_{em} = 0.726$ with transverse lines of sight separations of $2.94^{\prime}$ and $1.94^{\prime}$ from A \& B respectively. However, for this quasar only  G160M grating observations are available, which covers a few of the significant metal lines like {\CII}, {\OVI}, {\NV}, in addition to {\HI}. These lines suffer from different levels of contamination primarily from absorption associated with a sub-DLA at a higher redshift along the same line of sight \citep{2013MNRAS.433..178C}, rendering line measurements difficult. The third sightline is hence excluded from our main analysis, with details deferred to the Appendix \ref{Appendix A}.

\section{Spectral Analysis \& Ionization Modelling}

The angular separation of $\Delta \theta = 1.29^{\prime}$ between the sightlines A \& B corresponds to a transverse separation of $520$~kpc at the absorber redshift of $z = 0.399$. The redshift is defined based on the pixel showing maximum optical depth in the narrow {\CIII}~$977$~{\AA} profile along sightline A. The line measurements were carried out using both the integrated apparent optical depth (AOD) method of \citet{1991ApJ...379..245S} and by modelling the absorption lines using Voigt profiles. The AOD method relies on converting the flux profiles of unsaturated lines into apparent column densities which are then integrated over the velocity spread of the absorption. For saturated lines, this approach offers a lower limit on the column density. Voigt profile fits were performed on the detected lines using the VPFIT routine \citep[ver.10.4,][]{2014ascl.soft08015C}\footnote{The vpfit ver.10.4 is provided by the developers on https://people.ast.cam.ac.uk/~rfc/vpfit.html for public use.} after convolving the model profiles with the appropriate line spread functions\footnote{http://www.stsci.edu/hst/instrumentation/cos/performance/spectral-resolution}. For undetected lines, column density upper limits are derived by converting the $3\sigma$ equivalent width limit into column density assuming the linear relationship of the Curve of Growth. 

Photoionization equilibrium (PIE) models were generated using Cloudy  \citep[ver C17.01,][]{2017RMxAA..53..385F}, assuming the absorption to be from uniform density isothermal clouds of plane-parallel geometry. We adopt the solar relative elemental abundance pattern of \citet{2009ARA&A..47..481A}. The photoionization in the cloud is assumed to be regulated by the extragalactic background radiation (EBR) for the absorber redshifts as modelled by \citet{2019MNRAS.484.4174K} which incorporates the most recent measurements of the quasar luminosity function \citep{2009MNRAS.399.1755C,2013A&A...551A..29P} and star formation rate densities \citep{2015ApJ...805...33K}. To include ionizations resulting from the collisions of energetic free electrons with atoms and ions, we use the equilibrium and non-equilibrium collisional ionization (CIE, and NECI) models given by \citet{2007ApJS..168..213G}. The modelling results for the absorber at $z = 0.399$ along sightlines A \& B are summarized in Table \ref{tab:Summary_A&B}.  

\begin{table}
    \centering
    \caption{Line measurements for the absorber at z$_{abs} = 0.399$ towards Q~$0107-025$A (sightline A).}
    \label{tab:measure_A}
    \begin{threeparttable}
    \begin{tabular}{lcccr}
        \hline
        \multicolumn{5}{c}{Central component} \\
        \hline
        Line & W$_r$ (m\si{\angstrom}) & log N/{\cmsq} & b ({\kms}) & v (\kms) \\
        \hline
        H I 1215-917 & & 14.23 $\pm$ 0.05 & 22 $\pm$ 2 & 7 $\pm$ 4\\
        C III 977 & & 13.45 $\pm$ 0.16 & 11 $\pm$ 4 & 2 $\pm$ 1\\
        O VI 1031, 1037 & & 14.06 $\pm$ 0.08 & 33 $\pm$ 6 & 22 $\pm$ 3\\
        H I 1215 & 289 $\pm$ 13 & >14.0 & & [-50,70]\\
        H I 1025 & 87 $\pm$ 7 & 14.17 $\pm$ 0.02 & & [-50,25]\\
        H I 972 & 42 $\pm$ 11 & 14.28 $\pm$ 0.07 & & [-50,70]\\
        H I 949 & <30 & 14.34 $\pm$ 0.10 & & [-50,70]\\
        H I 937 & <50 & <15.0 & & [-50,70]\\
        H I 930 & <46 & <15.1 & & [-50,70]\\
        H I 926 & <27 & <15.1 & & [-50,70]\\
        H I 923 & <32 & <15.3 & & [-50,70]\\
        H I 920 & <40 & <15.6 & & [-50,70]\\
        H I 919 & <43 & <15.7 & & [-50,70]\\
        H I 918 & <43 & <15.8 & & [-50,70]\\
        H I 917 & <44 & <16.0 & & [-50,70]\\
        C III 977 & 95 $\pm$ 9 & 13.30 $\pm$ 0.03 & & [-30,60]\\
 		O VI 1031 & 108 $\pm$ 6 & 14.04 $\pm$ 0.02 & & [-20,75]\\
 		O VI 1037 & 66 $\pm$ 7 & 14.08 $\pm$ 0.02 & & [-20,75]\\
 		\hline
 		\multicolumn{5}{c}{$\sim$ -70 km/s component} \\
 		\hline
 		H I & & 13.52 $\pm$ 0.05 & 68 $\pm$ 18 & -64 $\pm$ 4\\
 		C III & & 12.85 $\pm$ 0.13 & 22 $\pm$ 11 & -64 $\pm$ 3\\
 		O VI & & 13.89 $\pm$ 0.11 & 38 $\pm$ 11 & -53 $\pm$ 3\\
 		H I 1215 & 102 $\pm$ 13 & 13.36 $\pm$ 0.06 & & [-130,-50]\\
        H I 1025 & 28 $\pm$ 7 & 13.62 $\pm$ 0.04 & & [-130,-50]\\
        C III 977 & 39 $\pm$ 8 & 12.84 $\pm$ 0.05 & & [-100,-40]\\
        O VI 1031 & 89 $\pm$ 6 & 13.93 $\pm$ 0.02 & & [-105,-15]\\
        O VI 1037 & 42 $\pm$ 7 & 13.86 $\pm$ 0.03 & & [-105,-15]\\
 		\hline
    \end{tabular}
    \begin{tablenotes}
    \small
    \textit{Note.} The columns show, from left to right: name of the absorption line with their rest wavelength mentioned, equivalent width of the absorption line profile with $1\sigma$ error, column density of the fitted Voigt profile with $1\sigma$ error, Doppler b-parameter of the fitted Voigt profile with $1\sigma$ error and the centroid or the velocity range of the profile. {\Lyb} is contaminated over $25 < v < 90$~{\kms}. The velocity range used for estimating the equivalent width and apparent column density are truncated before that.
    
    \end{tablenotes}
    \end{threeparttable}
\end{table}

In Fig. \ref{fig:combined_spectra} we compare certain key absorption lines along both sightlines centered on $z = 0.399$, the adopted rest-frame of the absorber complex. From the relative strengths of the lines, it is evident that sightline B is intercepting higher column density of neutral gas compared to sightline A, with significant differences in strength of low ionization ({\CII}, {\OII}, {\CIII}, {\OIII}, {\NIII}, {\SiIII}) absorption between the sightlines. However, the {\OVI} doublets have comparable strengths along both the lines-of-sight, implying that the {\OVI} is either tracing a gas phase that is more widely distributed than the patchier low ionization phase, or is produced in similar zones of ionization present along both sightlines. We now discuss the individual absorption systems and their ionization models in greater detail. 

\subsection{System towards Q$0107-025$A at $z_{abs} = 0.39949$}

\begin{figure}
\begin{center}
	\includegraphics[width=\columnwidth]{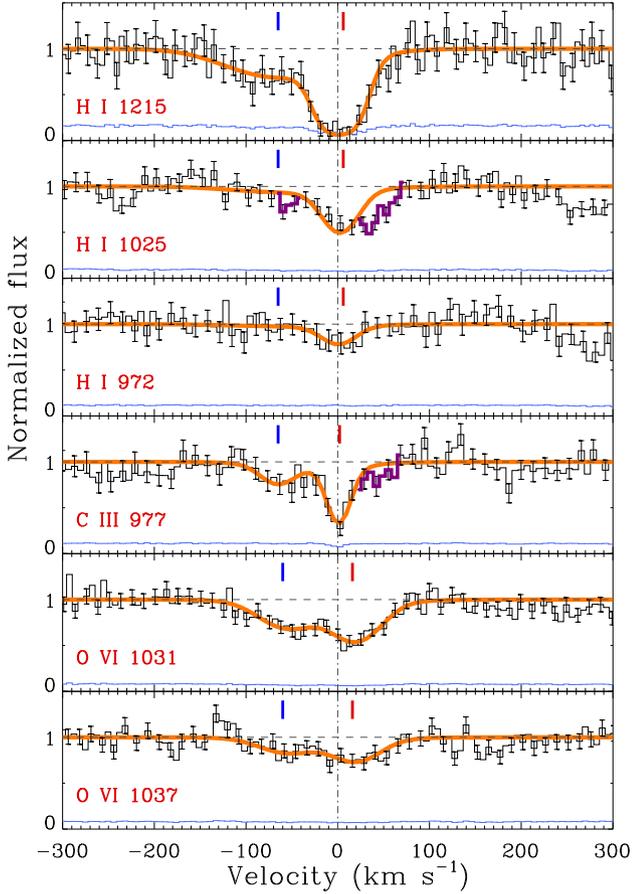}
    \caption{The continuum normalized absorption profiles of all lines detected for the $z_{abs} = 0.39949$ towards Q$0107-025$A. The $v = 0$~{\kms} in the X-axis corresponds to the redshift of the absorber. The best-fit Voigt profiles are also shown in each panel as \textit{orange} solid lines. The centroids of the components are indicated by the vertical tick marks. The {\Lyb} is contaminated from $25 \lesssim v \lesssim 90$~{\kms} (shown in \textit{purple}) where the absorption is inconsistent with {\Lya} and {\Lyg}. An $N_a(v)$ comparison of {\Lyb} with {\Lyg} shows that the narrow feature at $v \sim -60$~{\kms} of the {\Lyb} is contamination (shown in \textit{purple}), and is excluded from the {\Lyb} profile fit.}
    \label{fig:A_detections}
\end{center}
\end{figure}

\begin{figure*}
\begin{center}
	\includegraphics[width=1.0\textwidth]{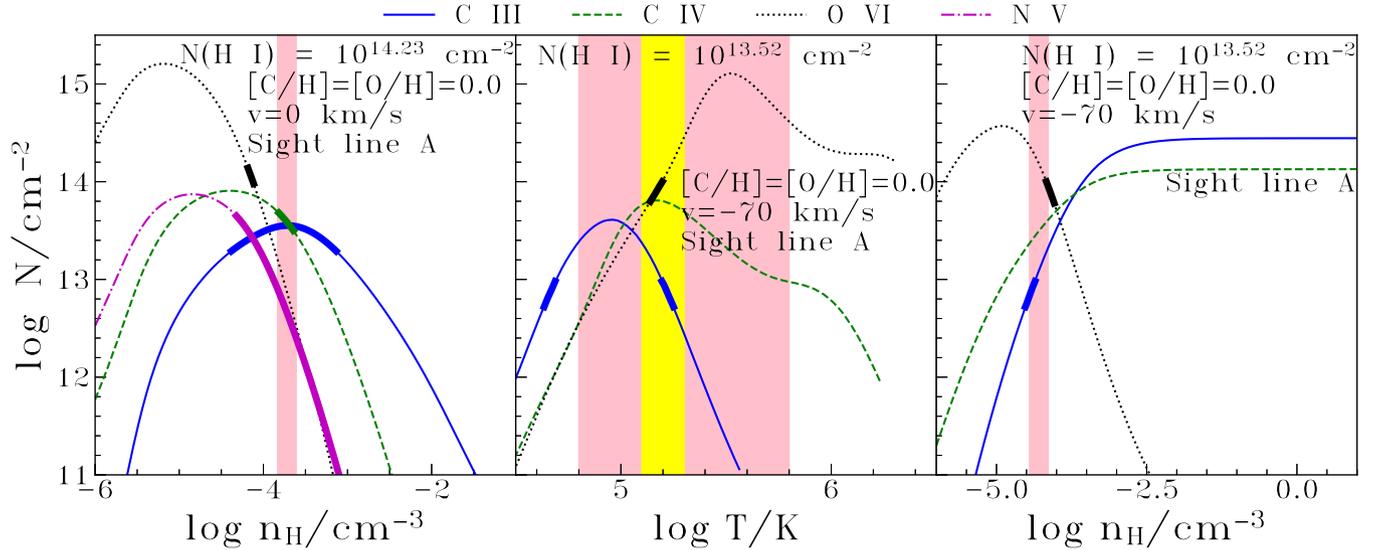}
    \caption{The three plots show the ionization model results for the absorber at $z_{abs} = 0.39949$ towards sight line A. The \textit{left} panel is the column density predictions from photoionization equilibrium models for different gas densities for the central component. The agreement between the observed column densities (shown in thick lines with $1\sigma$ error, upper limit of observed column density for {\NV} line) with the model predictions to within $1\sigma$ is indicated by the \textit{pink} band for solar metallicity. The column density of {\CIV} was taken from \citet{2010MNRAS.402.1273C}. The \textit{middle} panel shows the column density predictions from collisional ionization equilibrium models for different temperatures for the offset component at $\sim -70$~{\kms} at solar metallicty with thick lines indicating observed column densities with $1\sigma$ error. The model independent temperature of this cloud given by the different $b$-values of {\HI} and {\OVI} is shown by the \textit{pink} band. This $T = 10^{5.3}$~K at which the {\CIII} and {\OVI} are simultaneously explained is indicated by the yellow gap in the band. The \textit{right} panel shows hybrid model predictions for the offset component with thick lines being the observed column densities with $1\sigma$ error. Adopting the same metallicity as the central photoionized gas, the curve encompass the column density predictions for {\CIII} and {\OVI} by adopting mean temperature of $T = 1 \times 10^5$~K for the temperature range of $T = (0.5 - 1.5) \times 10^5$~K, for which the models are consistent with the data. In \textit{middle} and \textit{right} panels, predicted column density of {\CIV} is also shown.}
    \label{fig:Q0107025A_ion_modelling}
\end{center}    
\end{figure*}

The absorption at $z = 0.399$ towards sightline A is seen in {\Lya}, {\Lyb}, {\Lyg} and {\CIII}~$977$ at $\geq 3\sigma$ significance, in addition to the {\OVIdblt}. The absorption has two well resolved components at $v = 0$~{\kms} (central) and $v = -70$~{\kms} (offset). The system plot with detected lines is shown in Fig. \ref{fig:A_detections} and the corresponding line measurements are listed in Table \ref{tab:measure_A}. We determine the physical and chemical conditions in the two components separately by modelling them individually. The column density of {\HI} in the central component is constrained by simultaneous Voigt profile fit to the {\Lya}, {\Lyb} and {\Lyg} lines. The {\Lya} is saturated at the line core, whereas the {\Lyb} and {\Lyg} are unsaturated. The broad {\OVI} profile with $b(\OVI) > b(\HI) > b(\CIII)$ clearly suggests that the line of sight is probing a multi-phased medium, with bulk of the {\HI} possibly associated with a lower ionization gas phase traced by {\CIII}. The different $b$-values of {\HI} and {\CIII} solve for a temperature of $T = 2.4^{+0.6}_{-0.6} \times 10^4$~K for this phase. The inferred temperature is consistent with photoionized intergalactic/circumgalactic gas. The PIE model for this phase is shown in Fig. \ref{fig:Q0107025A_ion_modelling}(\textit{left panel}) where the column density for {\CIV} from \citet{2010MNRAS.402.1273C}, measured from $HST$/FOS G190H spectra, is also included. The creation and ionization energies of {\CIV} (48 eV and 64 eV) are less than half that of {\OVI} (113 eV and 138 eV). Multiple lines-of-sight observations often find different transverse scales for the absorption from {\CIV} and {\OVI} suggesting that the two ions could be tracing differently distributed gas phases \citep{1999ApJ...513..598L, 2007A&A...469...61L, 2020ApJ...900....9L}. Substantial fraction of the observed {\CIV} is likely to be from the same phase as {\CIII}. We, therefore, associate the {\CIV} with the {\CIII} phase rather than the {\OVI} phase. The observed column density ratio of {\CIII} to {\CIV} is recovered by the PIE models at $n_{\H} = 2.0 \times 10^{-4}$~{\cc}. At this density, the observed $N(\CIII)$ and $N(\CIV)$ require the carbon abundance to be [C/H] = 0. Such a PIE model predicts a total hydrogen column density of $\log~[N(\H)/\cmsq] \approx 17.6$, a gas pressure of $p/K \approx 4.8$~{\cc}~K, and a path length along the line of sight of $L \approx 0.7$~kpc. The metallicity estimated carries an uncertainty of $\approx 0.2$~dex from the cumulative uncertainties in the {\HI} and metal lines. The intrinsic shape of the QSO SED contributing to the ionizing background can introduce an additional uncertainty. Different EBR models generated using a range of observationally consistent QSO SEDs in KS19 yield metallicities that differ by $\approx 0.2$~dex (Acharya \& Khaire, in preparation). For solar [C/O], the {\OVI} predicted from this phase is a factor of 50 less, thus requiring a separate phase to explain its origin. 

Explaining the origin of {\OVI} in the central component through an additional phase would require information on the associated {\HI}. The $b(\OVI)$ sets the temperature in such a phase to $T \leq 1.4 \times 10^6$~K. At the upper limits of these temperatures the neutral hydrogen fraction $f_{\HI} \sim N(\HI)/N(\H) \sim 1.5 \times 10^{-7}$ \citep{2007ApJS..168..213G} is such that the associated {\HI} will be shallow, and thermally broadened. The kinematic superposition of such a weak component on the strong central {\HI} makes it undetectable. Thus, the metallicity or density of the {\OVI} phase remains undetermined, though the broad $b(\OVI)$ allows the presence of warm gas. 

The {\Lya}, {\CIII} and {\OVI} in the offset component at $v \sim -70$~{\kms} are detected at lower significance compared to the central component. The uncertainty in the fit parameters, especially in the $b$-values, reflect the weak nature of this component. The $b$-values of {\CIII} and {\OVI}, though different, are within $1\sigma$ of each other. The accompanying {\Lya} is significantly broader suggesting temperature influence in the line broadening. The $b(\HI) = 68~\pm~18$~{\kms} makes this a broad-{\Lya} (BLA) feature\footnote{By definition, a BLA is a {\Lya} line with $b(\H) \geq 40$~{\kms} corresponding to $T \gtrsim 10^5$~K, assuming pure thermal broadening}. The $b$-values of the BLA in combination with {\CIII} and {\OVI} show this component to be tracing warm gas with $T = (2.1_{-1.2}^{+1.2}) \times 10^5$~K (i.e., $\log~[T/(K)] = 5.31_{-0.09}^{+0.09}$). The temperature predicted from the line widths spans the range where ionization fractions of {\CIII} and {\OVI} peak under collisional ionization. Higher metallicities would lead to enhanced cooling of the collisionally ionized gas by means of metal line emissions \citep{1989A&A...215..147B}. This leads to a rapid decline in the temperature of the plasma with electron-ion recombination reactions lagging behind. The delayed recombination would result in the collisional ionization fractions departing from their equilibrium value, particularly for $T < 5 \times 10^6$~K. We therefore explore both the equilibrium and non-equilibrium collisional ionization models of \citet{2007ApJS..168..213G} to explain the observed column densities. 

As shown in Fig. \ref{fig:Q0107025A_ion_modelling} (\textit{middle panel}), the observed $N(\CIII)$ and $N(\OVI)$ can be simultaneously recovered from such a NECI model for [C/H] = [O/H] = 0.0 and $T \sim 1.6 \times 10^5$~K. The metallicity is identical to that estimated for the photoionized, low ionization gas in the central component. A more realistic collisional model for the offset component should also include the additional ionization brought about by the extragalactic background radiation. Implementing photoionization and non-equilibrium collisional ionization simultaneously is beyond the scope of our current work. Photoionization from the extragalactic background radiation can play a key role in reducing the cooling of the metal enriched gas \citep{2009MNRAS.393...99W,2011MNRAS.413..190T}. 

We instead resort to hybrid models using Cloudy by fixing the temperature of the plasma to $T \sim 1.6 \times 10^5$~K at which collisions of electrons with atoms and ions induce ionization, while simultaneously allowing for photoionization by the extragalactic background radiation. In Fig. \ref{fig:Q0107025A_ion_modelling}(\textit{right panel}) we show such a hybrid model. There is no single density or temperature at which {\CIII} and {\OVI} are simultaneously explained. Instead, the observed column densities of {\CIII} and {\OVI} are recovered to within their $1\sigma$ values for $T \sim (0.5 - 1.5) \times 10^5$~K and $n(\H) \sim (0.4 - 0.7) \times 10^{-4}$~{\cc} for [C/H] = [O/H] = $0$ (similar to the central component). Adopting a mean value of $T = 1.0 \times 10^5$~K (as shown in Fig.~\ref{fig:Q0107025A_ion_modelling}), we get the mean value of $n_{\H} \sim 0.6 \times 10^{-4}$~{\cc} from the hybrid model. The model yields a total hydrogen column density of $\log~[N(\H)/\cmsq] \approx 18.5$, a gas pressure of $p/K \approx 5.4$~{\cc}~K, and a path length along the line of sight of $L \approx 19.5$~kpc for this medium that is ionized by photons and collisions with energetic electrons. Such a model also predicts $\log [N(\CIV)/\cmsq] = 13.48$, indicating that there could be some contribution to the {\CIV} absorption seen in the low resolution FOS data from the offset component also. The ionization models are simplistic. Instead of the absorbing medium having an uniform temperature throughout, the line of sight could very well be probing slightly different temperature - density regions across narrow physical scales that are kinematically undistinguished at the resolution and sensitivity of the data. 

To summarize, for the absorption seen in along sightline A, the low and intermediate ionization metal lines and bulk of the {\HI} in the central component is tracing photoionized gas of solar metallicity and $T \sim 10^4$~K. Such a phase significantly under-produces the {\OVI} that is coincident in velocity. In the kinematically offset component, the ionization model independent gas temperature given by the line widths of {\HI} and {\OVI} favor collisional ionization. It is plausible that the {\OVI} in the offset component is part of the same warm ($T \sim 10^5$~K) collisionally ionized structure. 

\begin{figure*}
\begin{center}
	\includegraphics[width=1.0\textwidth]{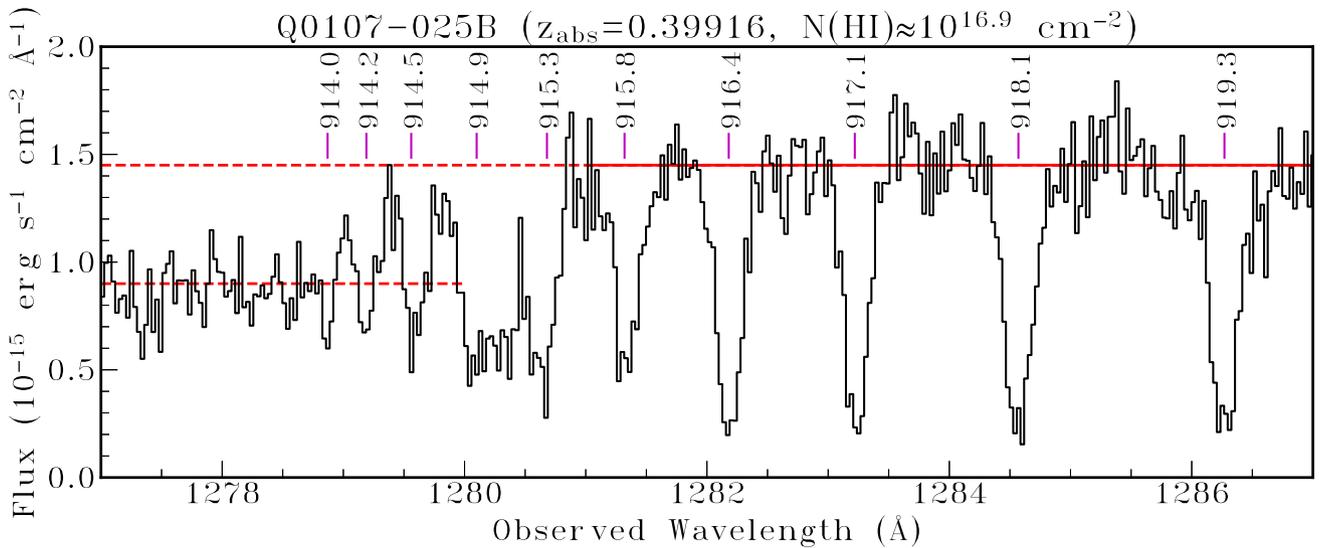}
    \caption{Higher order Lyman series lines and the partial Lyman break produced by the $z_{abs} = 0.39916$ absorber towards sightline B. The continuum level below and above the break are marked by the dashed and solid red lines respectively. The optical depth estimated from the partial Lyman break is 0.48$^{+0.11}_{-0.13}$, corresponding to $N(\HI) = 7.59^{+1.74}_{-2.10} \times 10^{16}$ {\cmsq}.}
    \label{fig:PB6291_LL}
\end{center}
\end{figure*}

\begin{table*}
    \centering
    \caption{Line measurements for the absorber at $z_{abs} = 0.399$ towards Q~$0107-025$B (sightline B).}
    \label{tab:measure_B}
    \begin{threeparttable}
    \begin{tabular}{lcccr}
        \hline
        Line & W$_r$ (m\si{\angstrom}) & log N/\cmsq & b (\kms) & v (\kms)\\
        \hline
        \multicolumn{5}{c}{2-COMPONENT {\HI} FIT} \\
        \hline
        H I 1215-915.8 & & 16.82 $\pm$ 0.06 & 25 $\pm$ 4 & -8 $\pm$ 2\\
         & & 14.84 $\pm$ 0.16 & 64 $\pm$ 9 & 2 $\pm$ 1\\
        C III 977& & 14.63$^{+0.27}_{-0.41}$ & 22 $\pm$ 1 & -7 $\pm$ 2\\
        O III 832.9 & & 15.23$^{+0.47}_{-0.42}$ & 22 $\pm$ 4 & -4 $\pm$ 3\\
        Si III 1206 & & 13.80$^{+0.25}_{-0.28}$ & 22 $\pm$ 2 & -11 $\pm$ 3\\
        \hline
        \multicolumn{5}{c}{3-COMPONENT {\HI} FIT} \\
        \hline
        H I 1215-915.8 & & 15.60 $\pm$ 0.05 & 14 $\pm$ 5 & -27 $\pm$ 4\\
         & & 16.87 $\pm$ 0.07 & 18 $\pm$ 3 & -3 $\pm$ 1\\
         & & 14.97 $\pm$ 0.15 & 58 $\pm$ 5 & 12 $\pm$ 2\\
        C III 977 & & 13.50 $\pm$ 0.14 & 14 $\pm$ 4 & -30\\
         & & 14.60 $\pm$ 0.46 & 16 $\pm$ 7 & 1\\
        O III 832.9 & & 14.21 $\pm$ 0.40 & 13 $\pm$ 5 & -27\\
         & & 15.03 $\pm$ 1.29 & 15 $\pm$ 1 & 6\\
        Si III 1206 & & 13.00 $\pm$ 0.25 & 8 $\pm$ 3 & -30\\
         & & 13.80 $\pm$ 0.36 & 14 $\pm$ 4 & 0\\
        \hline
        C II 903.9, 903.6, 1036 & & 13.70 $\pm$ 0.05 & 20 $\pm$ 4 & -2 $\pm$ 1\\
        O II 834.4, 833.3, 832.7 & & 14.08 $\pm$ 0.05 & 18 $\pm$ 0 & -2 $\pm$ 0\\
        N III 989 & & 13.78 $\pm$ 0.02 & 17 $\pm$ 9 & -10 $\pm$ 3\\
        O VI 1031, 1037 & & 14.36 $\pm$ 0.05 & 52 $\pm$ 4 & 2 $\pm$ 1\\
        H I 1215 & 789 $\pm$ 20 & >14.4 & & [-130,110]\\
        H I 1025 & 503 $\pm$ 12 & >15.2 & & [-130,110]\\
        H I 972 & 361 $\pm$ 10 & >15.5 & & [-130,110]\\
        H I 949 & 315 $\pm$ 10 & >15.8 & & [-130,110]\\
        H I 937 & 281 $\pm$ 17 & >15.9 & & [-130,110]\\
        H I 926 & 245 $\pm$ 10 & >16.2 & & [-130,110]\\
        H I 923 & 200 $\pm$ 8 & >16.3 & & [-90,50]\\
        H I 920 & 202 $\pm$ 9 & >16.5 & & [-90,50]\\
        H I 919 & 182 $\pm$ 11 & >16.5 & & [-90,50]\\
        H I 918 & 172 $\pm$ 11 & >16.6 & & [-90,50]\\
        H I 917 & 147 $\pm$ 11 & >16.6 & & [-90,50]\\
        H I 916.4 & 160 $\pm$ 11 & >16.8 & & [-90,50]\\
        H I 915.8 & 120 $\pm$ 11 & >16.6 & & [-60,50]\\
        C II 903.9 & 82 $\pm$ 9 & 13.64 $\pm$ 0.03 & & [-50,60]\\
        C II 903.6 & 57 $\pm$ 9 & 13.74 $\pm$ 0.04 & & [-50,60]\\
        C II 1036 & 44 $\pm$ 10 & 13.63 $\pm$ 0.04 & & [-50,60]\\
 		C III 977 & 255 $\pm$ 10 & >13.9 & & [-100,100]\\
 		N II 1083 & <43 & <13.6 & & [-100,100]\\
 		N II 915 & <25 & <13.4 & & [-57,2]\\
 	    N III 989 & 52 $\pm$ 12 & 13.83 $\pm$ 0.06 & & [-100,100]\\
 		N V 1238 & <94 & <13.7 & & [-100,100]\\
 		N V 1242 & <98 & <14.0 & & [-100,100]\\
 		O II 834.4 & < 110 & < 14.2 & & [-40, 45] \\
 		O II 833.3 & 49~$\pm$~12 & 14.05~$\pm$~0.12 &  & [-40, 45] \\
 		O III 832.9 & 200 $\pm$ 13 & >14.7 & & [-60,80]\\
 		O VI 1031 & 208 $\pm$ 12 & 14.34 $\pm$ 0.03 & & [-100,100]\\
 		O VI 1037 & 116 $\pm$ 13 & 14.35 $\pm$ 0.04 & & [-100,100]\\
 		Mg II 1239 & <94 & <16.5 & & [-100,100]\\
 		Si II 1260 & <135 & <13.0 & & [-100,100]\\
 		Si II 1193 & <82 & <13.1 & & [-100,100]\\
 		Si II 1190 & <82 & <13.4 & & [-100,100]\\
 		Si II 1020 & <55 & <14.6 & & [-100,100]\\
 		Si III 1206 & 269 $\pm$ 25 & >13.4 & & [-100,100]\\
 		S II 1253 & <96 & <14.8 & & [-100,100]\\
 		S II 1250 & <98 & <15.2 & & [-100,100]\\
 		S III 1190 & <82 & <14.5 & & [-100,100]\\
 		S III 1012 & <40 & <14.1 & & [-100,100]\\
 		S IV 1062 & <40 & <14.0 & & [-100,100]\\
 		S VI 944 & <38 & <13.4 & & [-100,100]\\
 		S VI 933 & <57 & <13.3 & & [-100,100]\\
 		Fe II 1260 & <137 & <14.6 & & [-100,100]\\
 		Fe II 1144 & <74 & <13.9 & & [-100,100]\\
 		Fe III 1122 & <69 & <14.1 & & [-100,100]\\
 		\hline
    \end{tabular}
    \begin{tablenotes}
    \small
    \textit{Note.} The columns show, from left to right: name of the absorption line with their rest wavelength mentioned, equivalent width of the absorption line profile with $1\sigma$ error, column density of the fitted Voigt profile with $1\sigma$ error, Doppler b-parameter of the fitted Voigt profile with $1\sigma$ error and the centroid or the velocity range of the profile. H I 930 has been excluded from the simultaneous Voigt profile fit as it is affected by an emission line.
    
    \end{tablenotes}
    \end{threeparttable}
\end{table*}

\subsection{System towards Q$0107-025$B at $z_{abs} = 0.39916$}

\begin{figure*}
\subfigure{\includegraphics[width=0.49\textwidth]{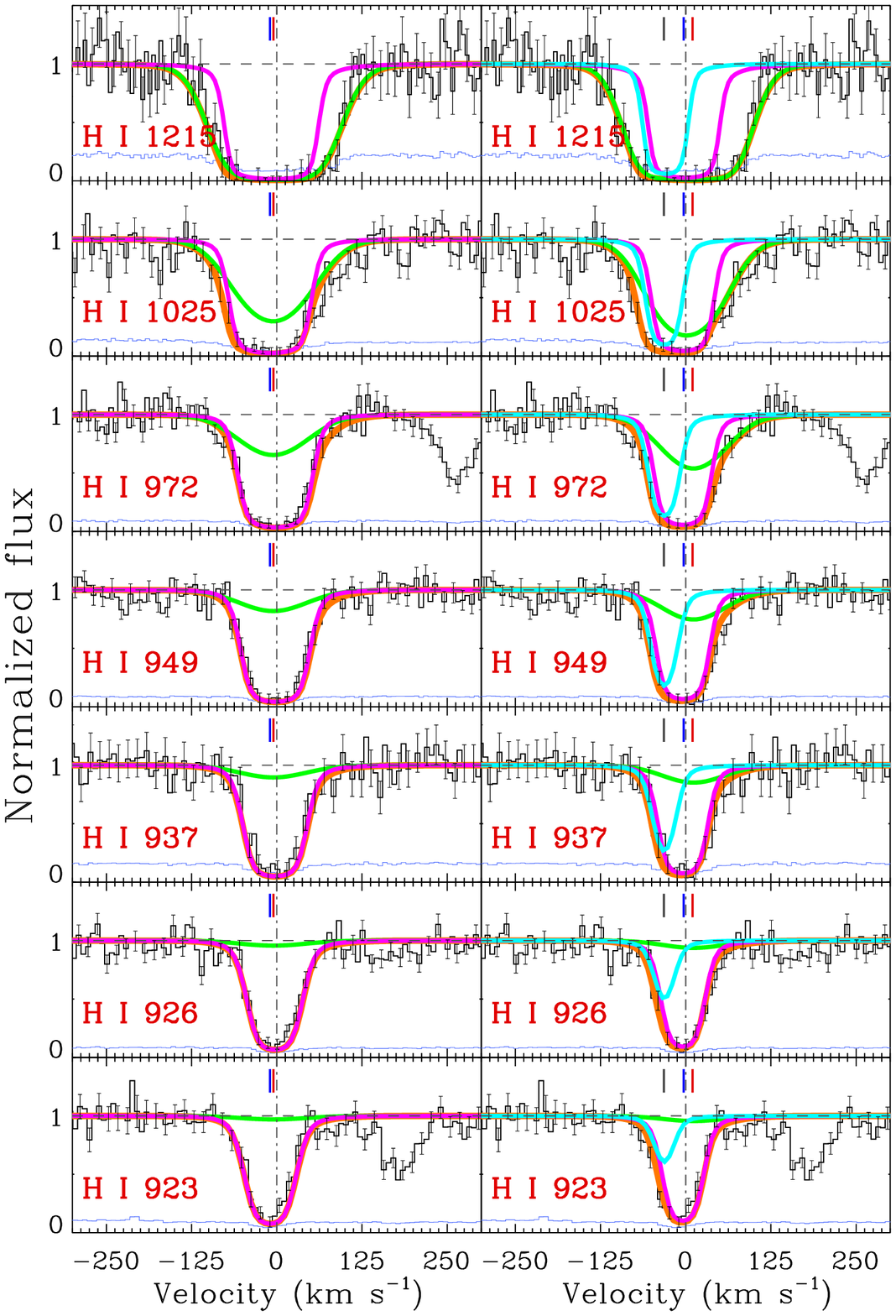}}
\hspace{0.01\textwidth}
\subfigure{\includegraphics[width=0.49\textwidth]{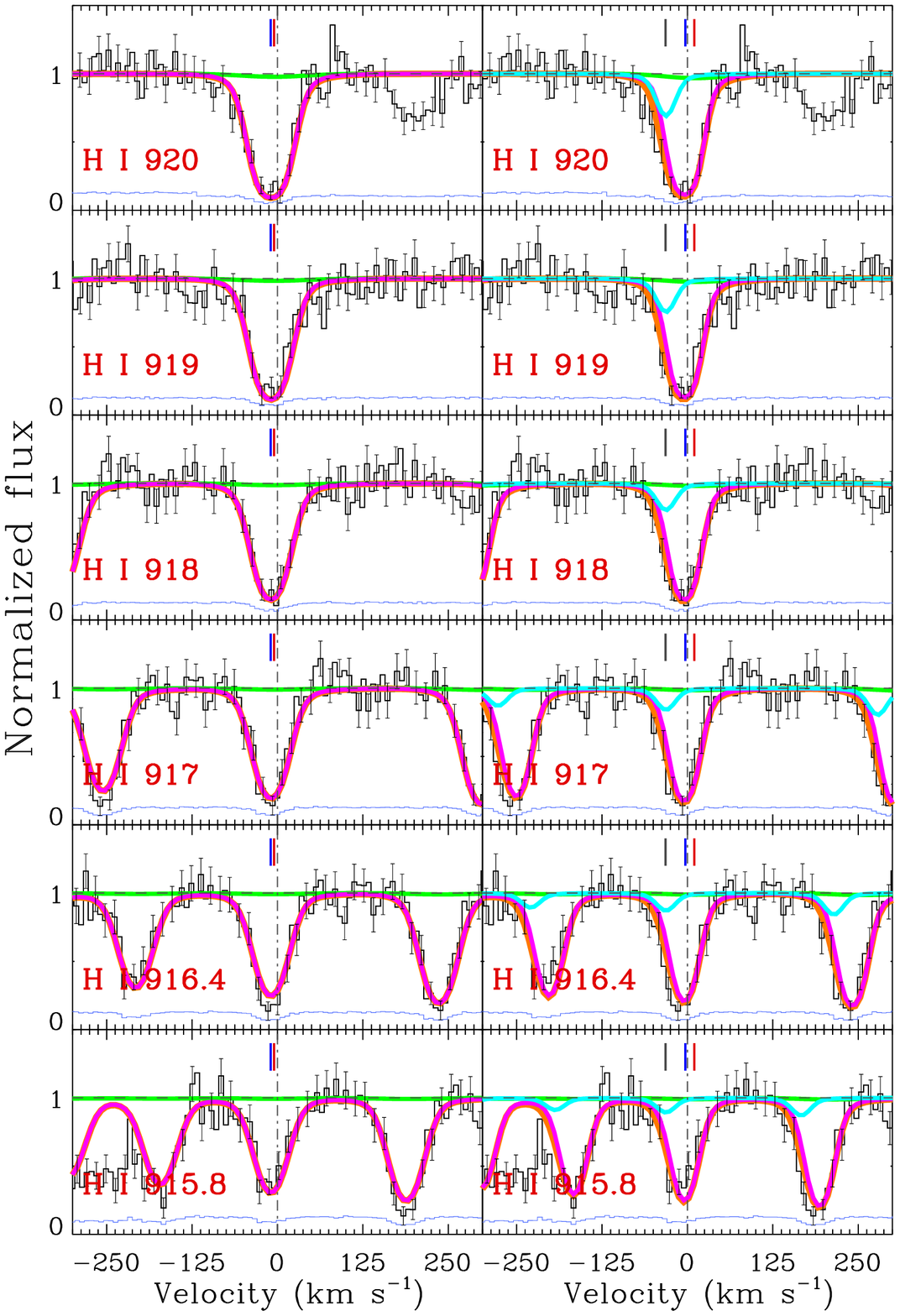}}
\caption{Continuum normalized Lyman series lines associated with the $z_{abs} = 0.39916$ absorber towards Q~$0107-025$B (sightline B). Here $v = 0$~{\kms} corresponds to the redshift of the absorber. The core {\HI} absorption is fitted with a single component profile and a two component profile, shown in adjacent panels. The component associated with the pLLS is indicated by the \textit{pink} Voigt profile, mainly constrained by the higher order Lyman lines. The addition of a narrow HI component at $v = -30$~{\kms} in the fitting can not explain the {\Lya}, {\Lyb} and {\Lyg} profiles. A BLA component is indeed required in both cases shown in \textit{green}. The BLA is related to the warm/hot phase traced by {\OVI}. The composite of the BLA and the stronger absorption is shown as the \textit{orange} profile. The centroids of the separate components are indicated by the vertical tick marks. The fitting parameters are listed in Table \ref{tab:measure_B}.}
\label{fig:B_detections_BLA}
\end{figure*}

\begin{figure*}
\subfigure{\includegraphics[width=0.49\textwidth]{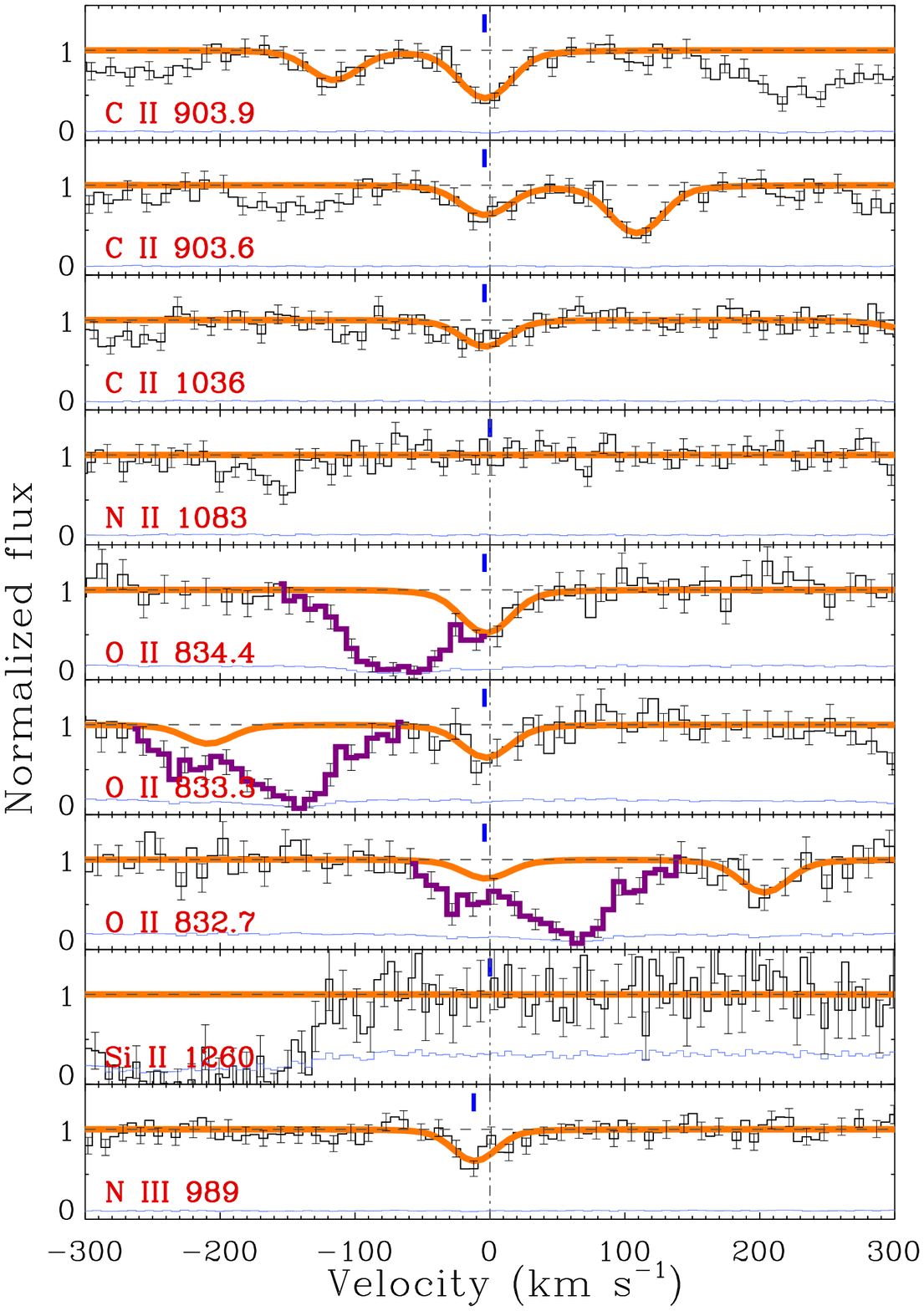}}
\hspace{0.01\textwidth}
\subfigure{\includegraphics[width=0.49\textwidth]{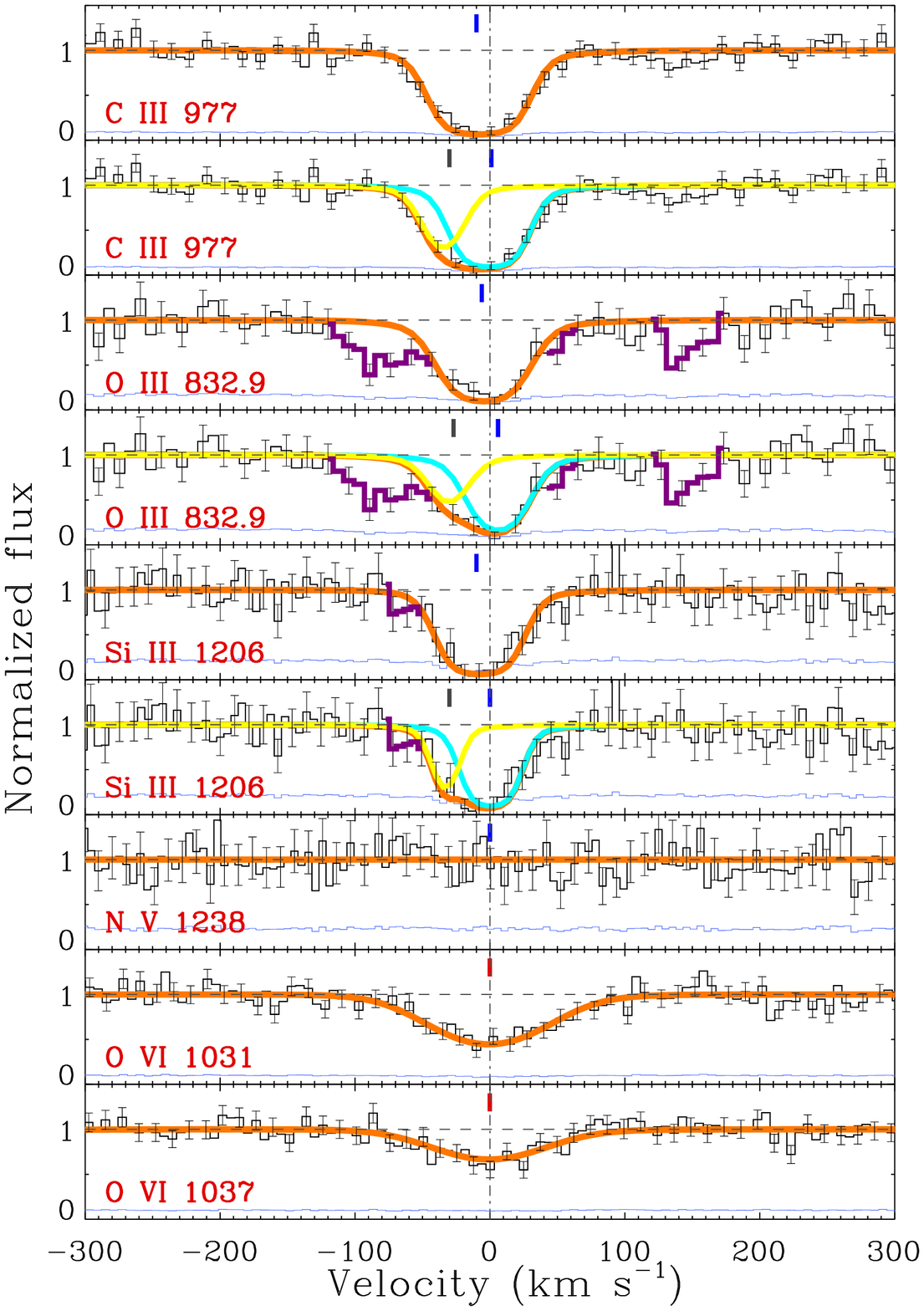}}
\caption{Figure similar to Fig.~\ref{fig:B_detections_BLA} shows the metal ions associated with the $z_{abs} = 0.39916$ absorber towards Q~$0107-025$B (sightline B), along with the Voigt profile models. The {\CIII}~977, {\OIII}~832.9, and {\SiIII}~1206 profiles are modelled using two possibilities, a single component, and two components. In the two component fits, the contribution from the individual components are also separately shown. The fitting results are summarized in Table \ref{tab:measure_B}. The part of the spectra which is contaminated is shown in \textit{purple}.}
\label{fig:B_detections_metals}
\end{figure*}

The absorber at z$_{abs}$=0.399 along the Q$0107-025$B sightline is partially optically thick in {\HI}. The optical depth estimated from the partial Lyman break is $\tau_{{912} {\AA}} = 0.48^{+0.11}_{-0.13}$ corresponding to $\log~[N(\HI)/\cmsq] = 16.88^{+0.09}_{-0.14}$, similar to that obtained from simultaneously fitting the Lyman series lines from {\HI}~$1215 - 915.8$~{\AA}. The uncertainty in optical depth was derived by varying the continuum slightly within the limits of the uncertainty in the flux values. The partial Lyman limit break is shown in Fig. \ref{fig:PB6291_LL}. Applying a single component Voigt profile simultaneously to the Lyman transitions results in a fit that does not explain the broad absorption wings in {\Lya} and {\Lyb} (see Fig. \ref{fig:B_detections_BLA}). The core component is primarily governed by the higher orders of the Lyman series which have narrow profiles, are less saturated, and therefore better constrained. The {\Lya}, and to a lesser extent {\Lyb}, suggest the presence of an additional component. The significance of this is explained and discussed later in a subsequent paragraph in this section. Concurrent with the {\HI} absorption are lines from {\CII}, {\OII}, {\CIII}, {\NIII}, {\OIII}, {\SiIII} and {\OVI} that are detected at $> 3\sigma$ (see Fig.~\ref{fig:B_detections_BLA} and Fig.~\ref{fig:B_detections_metals}). Prominent non-detections at this redshift include {\NII}, {\SiII}, {\NV} and {\SVI}. The velocity profiles of all prominent lines are shown in Fig.~\ref{fig:B_detections_BLA} and Fig.~\ref{fig:B_detections_metals} and the line measurements in Table \ref{tab:measure_B}. The {\OIII}~$832$~{\AA} line is contaminated from $-145$ to $-60$~{\kms}. As shown in Fig. \ref{fig:B_detections_metals}, a portion of the contamination could be {\OII}~$832.7$~{\AA} associated with the same absorber. The contamination is demonstrated by the model profile for {\OII}~$832.8$~{\AA} line obtained by simultaneously considering the {\OII}~$834.5$~{\AA} and {\OII}~$833.3$~{\AA} absorption. To account for the contamination from {\OII}~$832.8$~{\AA}, we de-weighted the pixels from $1164.80$~{\AA} to $1165.20$~{\AA} while fitting a Voigt profile to the {\OIII}~$832$~{\AA} line.

The {\CIII}~$977$~{\AA}, {\OIII}~$832$~{\AA}, and {\SiIII}~$1206$~{\AA} lines are strongly saturated as observed from lower AOD integrated column densities of these lines (lower by $0.7$~dex, $0.4$~dex and $0.4$~dex respectively) from their profile fit values. The line saturation prevents us from obtaining a unique Voigt profile solution for these lines. The uncertainties given by the automated fitting routine does not account for the line saturation or possible sub-component structure. We therefore explore the range of single component profile models that can adequately match the observed line profiles by varying the $b$-parameter and column density of these three lines over a permissible set of values. These intermediate ion lines can only be as wide as $b(\HI)$ and cannot be narrower than $b(\CII)$, if tracing the same gas phase. By fixing the $b$-value to these two extreme possibilities, the column density limits are arrived at by synthesizing profiles that explain the data. By taking the midpoint of these limits, the column densities are constrained to $\log [N(\CIII)/{\cmsq}] = 14.63^{+0.27}_{-0.41}$, $\log [N(\OIII)/{\cmsq}] = 15.23^{+0.47}_{-0.42}$, and $\log [N(\SiIII)/{\cmsq}] = 13.80^{+0.25}_{-0.28}$. The uncertainties refer to the range of feasible column densities these ions can take when having $b$-parameter between aforementioned limits. Also, the uncertainties are higher compared to the values obtained from a free profile fit to these saturated lines. 

We also attempt a two component model to C III, O III, and Si III lines with the component centroids fixed at $v \sim -30$ km/s and $v \sim 0$ km/s based on visual inspection. Such a model results in a fit in which the total column density is dominated by the $v \sim 0$ km/s component, with values of $\log [N(\CIII)/{\cmsq}] = 14.60 \pm 0.46$, $\log [N(\OIII)/{\cmsq}] = 15.03 \pm 1.29$, and $\log [N(\SiIII)/{\cmsq}] = 13.80 \pm 0.36$. These estimates are within 1-$\sigma$ of the column densities we adopt for the intermediate ionization lines using a single component fit. We therefore proceed with our adopted single component column density measurements in the ionization models, but consider the two component model while fitting the {\HI} absorption. To generate a two component fit to the Lyman series lines, we fixed the centroids of the components to $v = - 30$~{\kms}, and $v = 0$~{\kms}, since there is no distinct indication of the component positions in the observed {\HI} profiles. The higher order Lyman lines are entirely consistent with a single component, suggesting that the absorption is dominated by the $v \sim 0$~{\kms} component. Such a two component model shown in Fig.~\ref{fig:B_detections_BLA}, and Table \ref{tab:measure_B} also does not explain the excess absorption in {\Lya}, leading to a third component in {\HI} as mentioned later in this section.

\begin{figure*}
\begin{center}
	\includegraphics[width=1.0\textwidth]{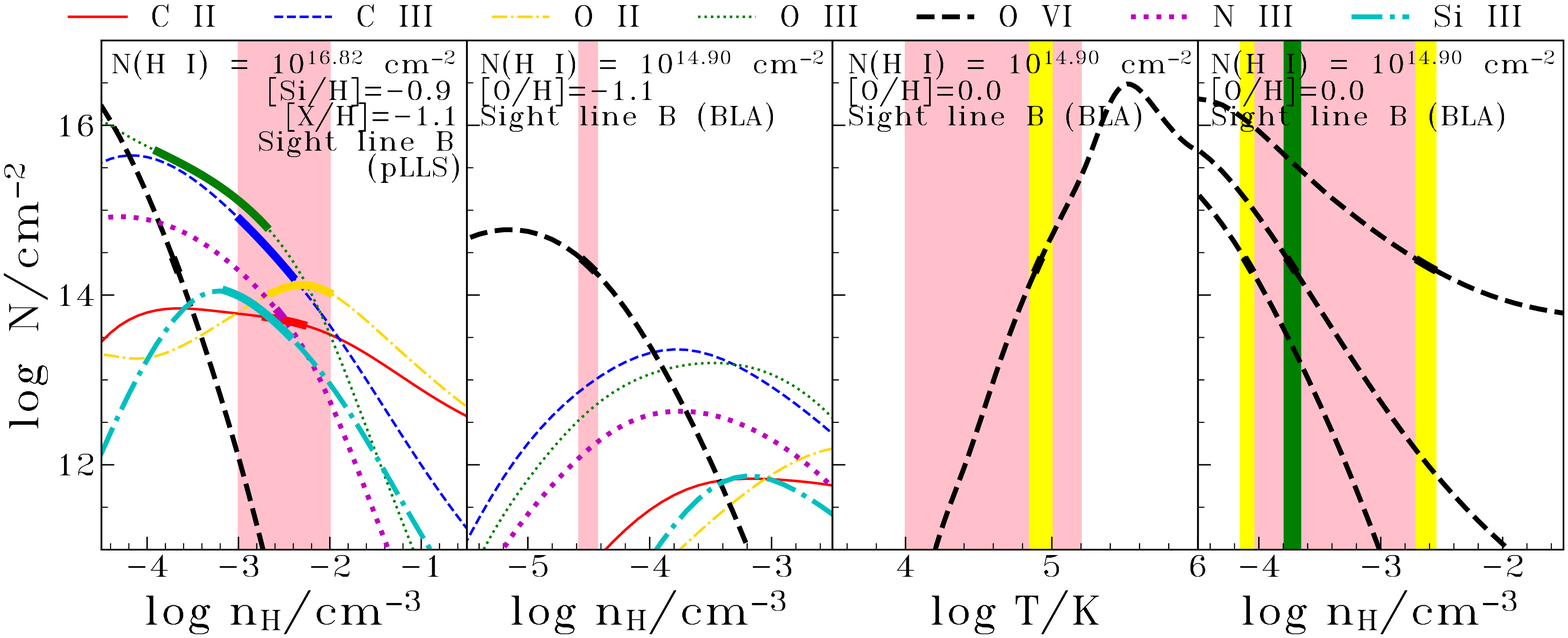}
    \caption{The plots show the ionization model results for the absorber at $z_{abs} = 0.39916$ towards Q~$0107-025$B (sightline B). The \textit{left} panel is the photoionization model predictions for the pLLS. The observed column densities of the metal ions except {\OVI} are simultaneously valid to within their $1\sigma$ uncertainty for the narrow density range indicated by the \textit{pink} region. The observed column densities with their $1\sigma$ uncertainties are indicated by the thick portion of the model curves. The origin of {\OVI} and the BLA from a higher photoionization is shown in the \textit{middle left} panel by adopting the same [O/H] as the low ionization pLLS phase. The {\HI} column density adopted is the mean of the column density values obtained for the BLA from a two component and a three component structure to the {\HI}. An alternative for the origin of {\OVI} - BLA absorption is collisional ionization. The \textit{middle right} panel shows collisional ionization predictions for {\OVI} in a radiatively cooling gas (non-equilibrium collisional ionization model, NECI). The observed column density is recovered at $T = 8.9 \times 10^4$~K for [O/H] = 0 (the band shown in \textit{yellow}). The shaded region (in pink) corresponds to the broad temperature range set by the $1\sigma$ lower and upper limits of $T = (0.1 - 1.4) \times 10^4$~K, from the combined $b$-values of BLA and {\OVI} for the two component and three component {\HI} models. The \textit{right} panel shows a hybrid of photoionization and collisional ionization equilibrium model for the {\OVI}-BLA phase. The \textit{top} curve corresponds to {\OVI} prediction for $T = 1.5 \times 10^5$~K, and the \textit{bottom} curve for $T = 0.1 \times 10^5$~K, which are the $1\sigma$ lower and upper temperature values from the two different BLA Voigt profile fit results. The density corresponding to the observed column density is indicated by the \textit{yellow} vertical shaded strip. The middle curve is the hybrid model prediction for the mean temperature of $T = 7.8 \times 10^4$~K, and the \textit{green} strip is the density at which $N(\OVI)$ is recovered for [O/H] = 0. Modelling results are summarized in Table \ref{tab:Summary_A&B}.}
    \label{fig:Q0107025B_ion_modelling}
\end{center}
\end{figure*}

Interestingly, the unsaturated {\OVIdblt} lines are a factor of $2.3$ broader than the low and intermediate ionization metal lines. This is an indication that the absorbing medium does not have a uniform temperature throughout. Though the low and intermediate ions overlap with {\OVI} in velocity space, the significantly broader profile of {\OVI} firmly establishes the multiphase nature of the absorption, independent of ionization models. The Cloudy PIE models for this absorber are shown in Fig. \ref{fig:Q0107025B_ion_modelling}. The observed $N(\CII)/N(\CIII) = -0.93^{+0.28}_{-0.42}$ and the $N(\OII)/N(\OIII) = -1.15^{+0.48}_{-0.42}$ are simultaneously explained at $n_{\H} = 2.5^{+7.5}_{-1.7} \times 10^{-3}$~{\cc} for [C/O] $= 0$ where the range is arrived at by taking into account the uncertainty in the C II and O II column densities. Here we have assumed that these two adjacent ionization stages of carbon and oxygen trace the same gas phase. In this phase, a metallicity of [O/H] = [C/H] $= -1.1~\pm~0.2$ is required to explain the observed column densities of {\OII}, {\OIII}, {\CII} and {\CIII}. The inferred metallicity carries an additional uncertainty of $\sim$ 0.2 dex due to the choice of UVB model, generated using one of a range of observationally consistent quasar SEDs in KS19 ionizing background radiation (Acharya \& Khaire in prep.). The combined uncertainty is more than the $\sim$ 0.1 dex uncertainty in metallicity if we adopt the two component profile model for the strong absorption in HI. The Si abundance has to be $0.2$~dex higher for in order to explain the observed $N(\SiIII)$ from the same phase. Such a PIE model for single component fit predicts a total hydrogen column density of $\log~[N(\H)/\cmsq] \approx 19.2$, a gas pressure of $p/K \approx 40$~{\cc}~K, and a line of sight thickness of $L \approx 2.2$~kpc. The {\OVI} from this phase is under-predicted by more than four orders of magnitude, indicating a separate phase. Evidence for such multiphase structure can also be gleaned from the broader $b$-parameter of {\OVI} compared to the low and intermediate ions, and from the profile fit to {\HI} as explained ahead.

As mentioned earlier, the single, and the two component profile models for the {\Lya} and {\Lyb} require the presence of an additional component. The inclusion of such a component with velocity centered on the {\OVI} significantly improves the fit to the {\Lya} and {\Lyb}. In the case where the {\HI} was previously fitted with a single narrow component, the second component has a $\log~N(\HI) = 14.84~{\pm}~0.16$ and $b(\H) = 64 ~{\pm}~9$~{\kms}, where the errors are statistical fit errors (see Fig. \ref{fig:B_detections_BLA}), ignoring the continuum fitting uncertainties. The $b$-value indicates that the second component is a BLA. The combined $b$-values of the BLA and {\OVI} give a temperature of $T = (8.9~\pm~5.6) \times 10^4$~K. In the alternative model for the {\HI}, where the core absorption was previously fitted with two narrow components, the third component is still a BLA, but with $\log~N(\HI) = 14.97~{\pm}~0.15$ and $b(\H) = 58 ~{\pm}~5$~{\kms}. The $b$ of the BLA, and the {\OVI} solves for $T = (4.3~\pm~3.3) \times 10^4$~K, which is within $1\sigma$ of the temperature estimated from the former two component fit.

We find that the {\OVI} along this sightline is consistent with both photoionization (PIE) in highly diffuse gas as well non-equilibrium collisional ionization (NECI). As the PIE models shown in Fig. \ref{fig:Q0107025B_ion_modelling} (\textit{second panel}) suggest, the observed {\OVI} can be recovered from a higher ionization phase provided [O/H] $\geq -1.4$. Assuming that the {\OVI} is from gas with similar metallicity as the lower ionization phase, adopting [O/H] $= -1.1$ we find that the observed $N(\OVI)$ is explained by the PIE models at $n_{\H} = 3.2 \times 10^{-5}$~{\cc}, with an equilibrium temperature of $T = 4 \times 10^4$~K. The photoionized {\OVI} is thus tracing gas that is more diffuse (two orders of magnitude lower in density) than the gas phase responsible for the low and intermediate ions, and the pLLS. Such a photoionized {\OVI} phase would contribute little C II, C III, or O III (see \textit{middle left panel} Fig.~\ref{fig:Q0107025B_ion_modelling}). The PIE model for this high ionization phase corresponds to a total hydrogen column density $N(\H) \approx 2.18 \times 10^{19}$~{\cmsq}, a gas pressure of $p/K \approx 1.28$~{\cc}~K$^{-1}$, and a line of sight thickness of $L \approx 221$~kpc. 

Collisional ionization in a cooling plasma is also a plausible mechanism for the {\OVI} to be produced, even at $T \lesssim 10^5$~K implied by the line widths. The column density prediction from the NECI models of \citet{2007ApJS..168..213G} for gas at solar metallicity is shown in Fig. \ref{fig:Q0107025B_ion_modelling} (\textit{third panel}). Such a model shows that for [O/H] $= 0$, the observed $N(\OVI)$ is explained at $T = 8.3 \times 10^4$~K ($\log [T/K] = 4.92$). The hybrid models involving both photoionization and collisional ionization were generated for  $T = 0.1 \times 10^5$~K and $T = 1.5 \times 10^5$~K  set by the $1\sigma$ range for the temperature from the two and three component {\HI} fitting models of Table \ref{tab:measure_B} (\textit{fourth panel of Fig. \ref{fig:Q0107025B_ion_modelling}}). This is a conservative range on the temperature that the BLA - {\OVI} bearing gas can have. Within these limits, the hybrid models are consistent with the observed $N(\OVI)$ for densities in the range of $n_{\H} = (0.1 - 2.5) \times 10^{-3}$~{\cc}. In these models, a metallicity of solar, similar to the metallicity determined for the BLA - {\OVI} gas along sightline A, is assumed. The [O/H] in the {\OVI} phase along sightline B cannot be constrained to a unique value due to the large temperature uncertainty from the inadequately constrained BLA $b$-parameter. A hybrid model with the $1\sigma$ lower limit on temperature of $T = 10^4$~K puts a conservative lower limit of [O/H] $\gtrsim -1.5$ while reproducing the observed $N(\OVI)$ at $n_{\H} = 10^{-5.2}~{\cc}$ at which the ionization fraction of {\OVI} peaks. In Table \ref{tab:Summary_A&B}, we summarize the ionization modelling results, where we have listed the NECI and hybrid model parameters, with uncertainties, for the two possible temperatures predicted for the BLA-{\OVI} phase from the two component and three component profile fits to the {\HI} lines.

\section{The Galaxy Environment \& Origins of Absorption}

\begin{figure*} 
    \centering \includegraphics[scale=0.7]{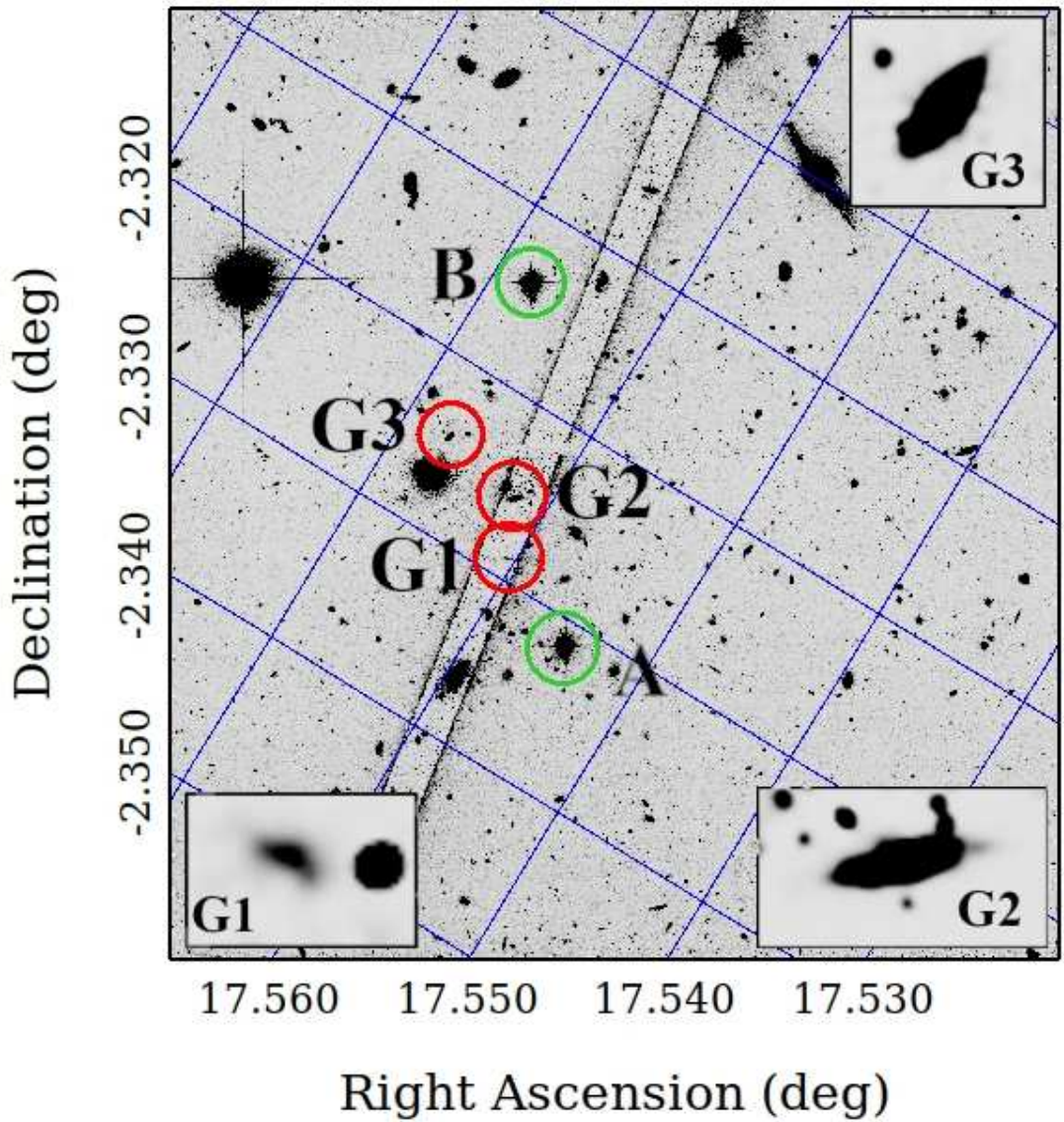}
    \caption{An $HST$/ACS image encompassing sightline A, sightline B and three of the nearest galaxies coincident in redshift with the absorbers. The diagonal lines are artifacts of the instrument. The grids (shown in blue) are lines of constant RA, and Dec. The closest galaxies to either sight lines are G1 ($i = 35.37~\pm~35.81^{\circ}$, $\alpha = -11.13~\pm~90.00^{\circ}$), G2 ($i = 72.5~\pm~3.3^{\circ}$, $\alpha = -46.60~\pm~11.87^{\circ}$) and G3 ($i = 81.84~\pm~1.07^{\circ}$, $\alpha = -8.88~\pm~0.79^{\circ}$). Quasars A and B are marked with green circles while G1, G2 and G3 are marked with red circles. The third sightline with ($\alpha, \delta$) = ($17.5583^{\circ}, -2.2828^{\circ}$), discussed in Appendix~\ref{Appendix A} is outside this ACS field. The inset image contains the zoomed images of these galaxies. The axes show the Right Ascension (in degrees) and Declination (in degrees).}
    \label{fig:HST_image}
\end{figure*}

 We extracted information on galaxies in the extended environment near to the absorbers from the \citet{2014yCat..74372017T} catalogue, which is a composite multi-object spectroscopic survey of galaxies using VLT/VIMOS, Keck/DEIMOS and Gemini/GMOS in the foreground fields of quasars (see Table~\ref{tab:Galaxy}). The details of those observations are summarized in Sec 3 and Table 4 of \citet{2014yCat..74372017T}. The spectrum from the VIMOS ($R \equiv \lambda/\Delta\lambda \approx 200$) and DEIMOS ($R \equiv \lambda/\Delta\lambda \approx 5000$) has a wavelength coverage of~$5500 - 9500$~{\AA} and~$6400 - 9100$~{\AA}, respectively and for GMOS ($R \equiv \lambda/\Delta\lambda \approx 640$), R400 grating centred on a wavelength of $7000$~{\AA} is used. We supplement these with spectroscopic data from VLT/MUSE, GMOS, and VIMOS for G1, G2, and G3 respectively (see Table~\ref{tab:Galaxy}) (Beckett {\etal}, in preparation), which are galaxies closest to the absorber in impact parameter. The MUSE observation of G1 is in the public archive (Program ID : 094.A-0131, PI: Joop Schaye). The MUSE has the wavelength range of $4800 - 9300$~{\AA} with resolving power of $1740 - 3590$. The projected separations to all these galaxies along with their luminosities, virial radii, stellar and halo masses, and star formation rates are listed in Table \ref{tab:Galaxy}. A brief overview of how these parameters are estimated is given in tablenotes of Table~\ref{tab:Galaxy}. An $HST$/ACS image of the field is shown in Fig. \ref{fig:HST_image}, that includes these three nearest galaxies. The medium resolution spectra of these galaxies are shown in Fig. \ref{fig:spectra_G1_G2_G3}.

\begin{figure*}
\begin{center}
  \includegraphics[width=\textwidth]{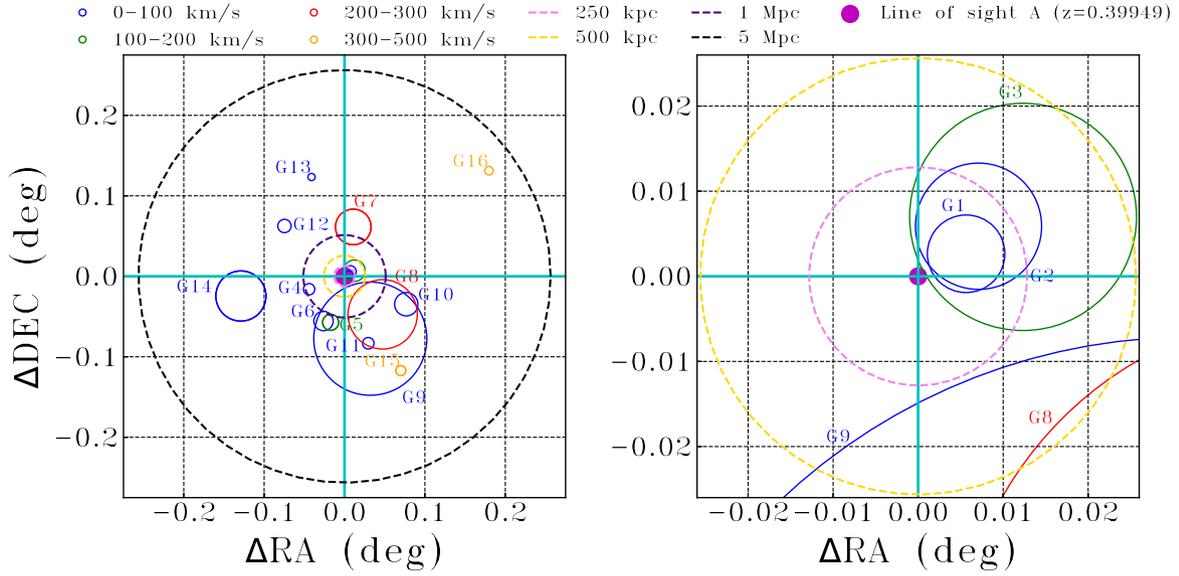}
\caption{Galaxies in the extended RA-Dec plane around the $z_{abs} = 0.39949$ towards Q~$0107-025$A (sight line A), with (0,0) corresponding to the quasar line of sight. Galaxies are numbered in increasing order of impact parameter with G1 corresponding to the nearest. Impact parameters of 250 kpc, 500 kpc, 1 Mpc and 5 Mpc are represented by the \textit{pink, yellow, indigo,} and \textit{black dashed} circles respectively in the \textit{top} panel. The galaxy's systemic velocity relative to the absorber redshift is given by the color-coding. The \textit{right} panel is a zoom-in to $500$~kpc of projected separation with the galaxies represented by open circles whose radii correspond to the $R_{vir}$ as determined in Table \ref{tab:Galaxy}.}
    \label{fig:Gal_A_0}
\end{center}
\end{figure*}

\begin{figure*}
\begin{center}
        \includegraphics[width=\textwidth]{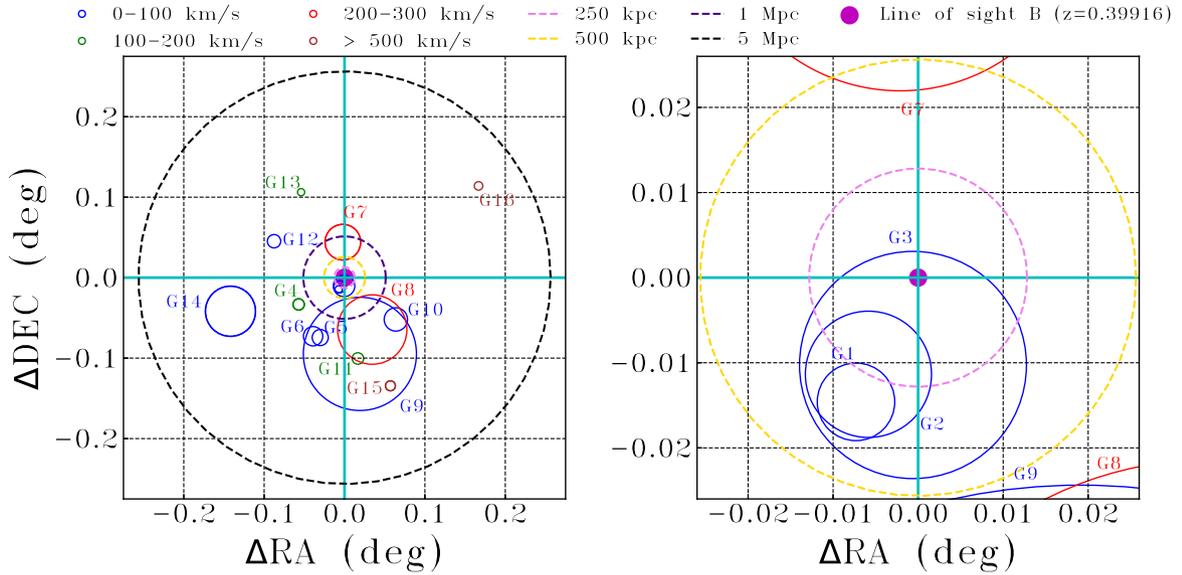}
    \caption{Same as Fig.~\ref{fig:Gal_A_0} but for sightline B.}
    \label{fig:Gal_B}
\end{center}
\end{figure*}

\begin{figure*}
\begin{center}
  \begin{subfigure}{
  \includegraphics[clip=true,trim={0cm 0cm 0cm 0cm},scale=0.4]{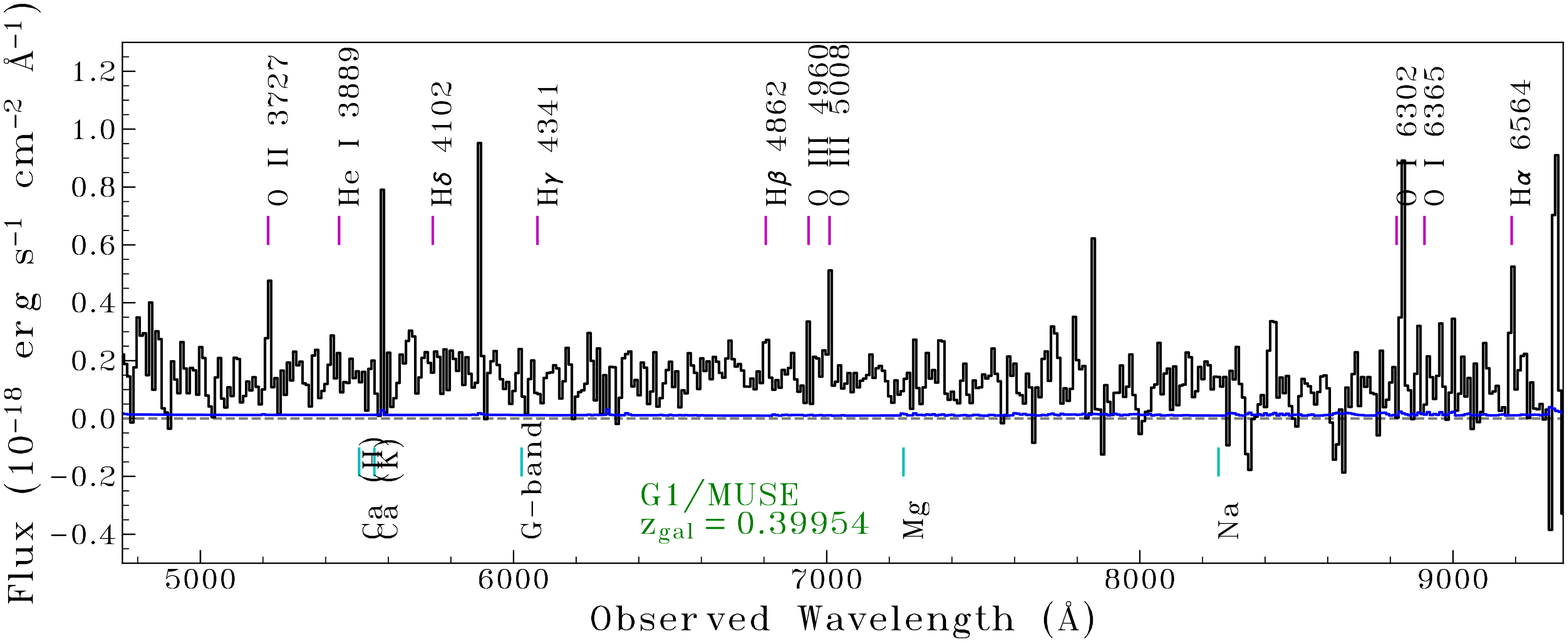}}
  \end{subfigure}
  ~
  \begin{subfigure}{
  \includegraphics[clip=true,trim={0cm 0cm 0cm 0cm},scale=0.4]{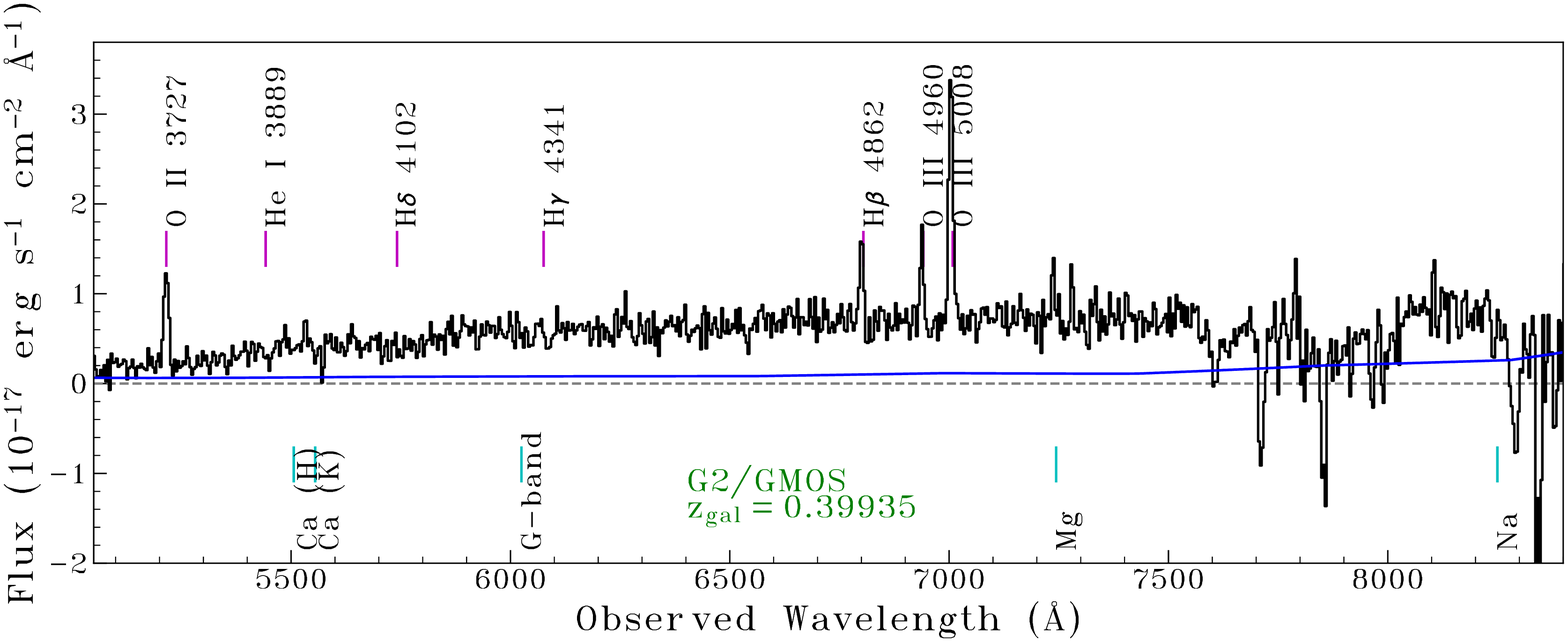}}
   \end{subfigure}
  ~
  \begin{subfigure}{
 \includegraphics[clip=true,trim={0cm 0cm 0cm 0cm},scale=0.4]{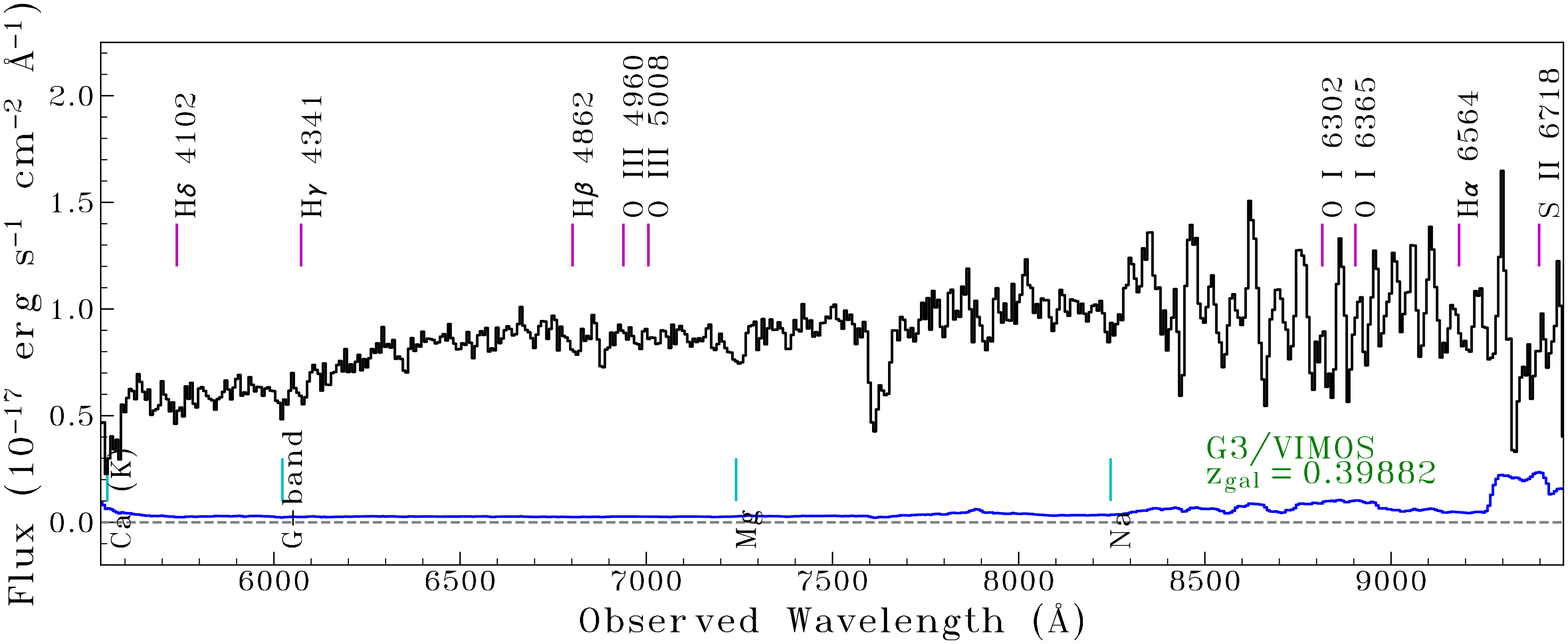}}
   \end{subfigure}
   
\caption{The panels show the MUSE/MOS spectra of the galaxies G1 (top), G2 (middle), and G3 (bottom) that are closest to the absorber at $z = 0.399$ in projected separation. The 1$\sigma$ error spectrum is plotted in blue, and the zero flux level is indicated by the \textit{grey} dashed line. Expected locations of prominent nebular emission lines are labeled on top of each spectrum, with prominent absorption features labelled underneath the spectrum.}
\label{fig:spectra_G1_G2_G3}
\end{center}
\end{figure*}

 Within a projected separation of $\rho < 2$~Mpc and a velocity offset of $|\Delta v| < 250$~{\kms} between the galaxy's systematic redshift and $z = 0.399$, 12 galaxies are identified with stellar masses in the range of $M_* \sim 10^{8.7} - 10^{11.5}$~M$_\odot$. The radial velocity and projected separation cut-offs correspond to the values that are typical of virialized galaxy concentrations such as groups and poor clusters \citep[e.g.,][]{1999fsu..conf..135B}. Four more galaxies are identified within a wider projected separation of $5$~Mpc. The distribution of these galaxies towards both sightlines are shown in Fig. \ref{fig:Gal_A_0} and Fig. \ref{fig:Gal_B}. Three of these galaxies (G1, G2, and G3) are within $300$~kpc of both sightline, and the remaining at $\gtrsim 1$~Mpc. The presence of several galaxies within narrow projected positions and radial velocities indicates that the lines-of-sight are probing different regions of a galaxy over-density such as a group or a cluster, with the absorption possibly arising in the CGM or the intragroup/cluster medium, consistent with what has been found for optically thick gas \citep{2017ApJ...846L...8M,2019MNRAS.485...30M,2019MNRAS.488.5327P,2020MNRAS.497..498C}. 

Galaxy G1 is the closest in projected separation ($\rho \approx 121$~kpc) to sightline A, and galaxy G3 ($\rho \approx 200$~kpc) to sightline B. The stellar mass of $M^* \sim 10^{8.7}$~M$_\odot$ and the $L \sim 0.003L^*$ of galaxy G1 resembles the higher mass end of the local dwarf galaxy population \citep[e.g.,][]{2008ApJ...686.1030O,2017A&A...606A.115H}, whereas G2 and G3 with $M^* \sim 10^{10}, 10^{10.7}$~M$_\odot$ have masses comparable to Milky Way \citep[$M^*_{MW} = 10^{10.8}$~M$_\odot$,][]{2015ApJ...806...96L}, but with sub-$L^*$ luminosities. Sub-$L^*$ (0.25 $\leq$ L/L$^*$ $\leq$ 0.76) galaxies are found to be associated with the optically thick absorbers \citep[e.g.,][]{2008AJ....135..922K,2009ApJ...694..734L,2010ApJ...714.1521C,2010ApJ...711..533K,2010ApJ...717..289S}. Galaxy G3 is quiescent with a specific star formation rate of, sSFR $< 2 \times 10^{-12}$~yr$^{-1}$ (SFR $< 0.1$~M$_{\odot}$~yr$^{-1}$) \citep{2014ApJ...783L..30U,2017ApJ...838...19W}, whereas galaxy G2 is forming stars at sSFR of $\sim 3 \times 10^{-10}$~yr$^{-1}$ (SFR $\sim 3$~M$_{\odot}$~yr$^{-1}$). Galaxies G1, G2, and G3 are at normalized transverse separations of $\rho/R_{vir} \sim 1.4,~1.2,~1.1$ with respect to sightline A, and $\rho/R_{vir} \sim 3.6,~1.7,~0.8$ with sightline B. Conventionally, gas within $R_{vir}$ is considered as belonging to the halo and those at larger impact parameters have a higher probability of being unbound to that galaxy, and likely related to the surrounding large-scale structures. The virial radius however provides only a characteristic value for the halo size. Environmental influences on the distribution of gas are known to extend beyond $R_{vir}$, especially when the galaxies are part of dense groups or clusters \citep[e.g.,][]{1999ApJ...524L..19M,2008MNRAS.383..593M,2013MNRAS.430.3017B}. Given that the galaxies close to the lines of sight are low- and intermediate-mass systems, the absorbers could be tracing interstellar gas displaced through stellar/AGN feedback or tidal interactions even when the impact parameters are not strictly within the virial halo. We now proceed to consider these possibilities. 

Due to impact parameters similar to virial radii of G2 and G3 with respect to sightline A and B, the sightlines could potentially be probing gas that has originated in either or both of these galaxies. The CGM of even low-luminosity galaxies $< 0.1L^*$ are known to possess substantial mass in the form of relatively cool gas ($T \sim 10^4 - 10^5$~K) in the range of $M_{cool} \sim 10^7 - 10^8$~M$_\odot$ \citep{2013ApJ...763..148S}. The circumgalactic gas is a heterogeneous population comprising of gas accreted from the IGM, outflows fueled by star-formation or nuclear activity, and interstellar gas of satellite galaxies displaced in tidal interactions \citep{2017ARA&A..55..389T}. The metallicity of the gas is a means to distinguish these different processes, with high-metallicity absorbers originating in metal-rich outflows \citep[e.g.,][]{2011Sci...334..952T,2015ApJ...811..132M}, and low-metallicity absorbers tracing gas getting accreted from the intergalactic medium \citep[e.g.,][]{2011ApJ...743..207R,2013Sci...341...50B}. Another discriminating factor is the azimuthal angle, defined as the angle between the QSO line-of-sight and the projected major axis of the galaxy hosting the absorber. Galaxy-quasar pair studies find biconical outflows preferentially directed along the galaxy minor axis with significantly strong warm absorption appearing along extensions closer to the minor axis \citep{2011ApJ...743...10B,2019MNRAS.490.4368S}. In contrast, accretion of cold streams of gas happens predominantly along the galaxy major axis \citep{2012ApJ...760L...7K,2015ApJ...812...83N,2020MNRAS.492.4576Z,2020ApJ...888...14H}. 

Relative to G2, sightlines A and B are aligned closer to the galaxy's projected minor axis at opposite sides with azimuthal angles of $\Phi \sim 86^{\circ}$ and $\Phi \sim 72^{\circ}$ respectively (see Fig.~\ref{fig:HST_image}). In comparison, the sightlines are at azimuthal orientations of $\Phi \sim 68^{\circ}$ and $\Phi \sim 3^{\circ}$ (aligned with projected major axis) with respect to G3. Thus, from the point of view of tracing metal-rich outflows, galaxy G2 is favorably oriented relative to the sightlines. Moreover, G2 has a higher current star-formation rate marked by the prominent emission lines in its spectrum whereas G3 is absorption line dominated (see Fig. \ref{fig:spectra_G1_G2_G3}). For characteristic outflow velocities of $\lesssim 500$~{\kms}, the time-scale for transporting metals to distances of $\sim 200$~kpc (average impact parameter of either sightline to galaxy G2) is of the order of $\gtrsim 600$~Myr \citep[e.g.,][]{2001ApJ...554.1021H,2009ApJ...692..187W}, which is an order of magnitude longer than the timescale of active outflows dictated by short-lived massive stars. The twin lines of sight are likely to be therefore probing gas enriched by biconical flows from past rather than on-going star formation in G2. The present SFR could be the culmination of the galaxy's past peak in star-formation.

The partial Lyman limit column of {\HI} along sightline B is consistent with enriched gas from G2 moving into the circumgalactic zone of galaxy G3 ($\rho \sim 0.8R_{vir}$ relative to G3), whereas the low column density gas along sightline A can be tracing either intragroup gas or the outskirts of the gaseous halo of G2 itself ($\rho \sim 1.2R_{vir}$). The metals dispersed from G2 can mix with the ambient neutral gas through diffusion or turbulence. They can also remain confined to patchy zones that are either transient \citep{2007MNRAS.379.1169S} or pressure bound within the hot CGM or the group environment. The characteristic temperature and density of $T \sim 1.5 \times 10^6$~K and $n_{\H} \sim 5 \times 10^{-5}$~{\cc} for the hot galactic coronae (assuming Milky Way halo mass) \citep{2017ApJ...835...52F} are adequate to keep the low ionization cloud with $T \sim 1.6 \times 10^4$~K and $n_{\H} \sim 2.5 \times 10^{-3}$~{\cc} in pressure equilibrium. This view is also consistent with the {\OVI} tracing transition temperature gas at the interface of the pLLS cloud and the galaxy's hot corona.

Since the abundance levels are inferred based on column densities integrated along a sightline, a larger column of metal-poor intergalactic gas along sightline B would naturally lead to an inferred metal concentration that is diluted in comparison to the lower {\HI} column density probed by sightline A. The probability of finding such a larger column of {\HI} gas along sightline B is higher as it penetrates close to galaxy G3 ($\rho \sim 0.8R_{vir}$), consistent with the anti-correlation trend seen in absorber-galaxy surveys between the distribution of {\HI} column density and impact parameter to the nearest galaxy \citep{1990ApJ...357..321L,1994AJ....108.2046S,2008AJ....135..922K,2008ApJ...687..745C,2010ApJ...714.1521C,2011ApJ...743...10B}. 

It has also been found that in dense environments, cool gas has a wider spread than around isolated galaxies \citep[e.g.,][]{2011ApJ...743...10B}. \citet{2018ApJ...869..153N} found significantly larger column densities of gas in pixels offset from galaxy systemic redshifts in group environments compared to {\MgII} absorption associated with field galaxies. They also found cool gas in groups having kinematics similar to that of outflows along the minor axis of galaxies. Together, it implies that group environments have substantial covering fraction of low-ionization gas mixed with metals mainly contributed by outflows and tidal interactions with member galaxies. The higher covering fraction increases the odds of multiple lines of sight intercepting such gas, compared to their smaller extensions around isolated galaxies \citep[see][]{2020arXiv200914219D}.  
An alternative explanation for the pLLS is for sightline B to be piercing an inflowing cold stream of low-metallicity gas from the intragroup medium into galaxy G3. The impact parameter of $\rho/R_{vir} \sim 0.8$ with G3, the inferred metallicity of $\sim 1/10$-th solar, and the alignment of the sightline closer to the galaxy's major axis are all consistent with this alternate scenario.  

In contrast to the low ionization gas, the {\OVI} along both the lines of sight has comparable column densities of $\log N \sim 14.1 - 14.3$, broad $b$-values of $\sim 33 - 52$~{\kms} and line of sight velocity separation of $|\Delta v| \approx 64$~{\kms}, consistent with tracing warm gas that is either part of a uniform large-scale medium such as the warm CGM, or restricted to a much narrower thickness transition temperature zone resulting in similar column densities. It is well known from observations and simulations that as many as $50$\% of the {\OVI} absorbers at $z \lesssim 1$ are tracers of collisionally ionized plasma with temperatures of $T \sim 10^5$~K \citep{2008ApJS..177...39T,2011ApJ...731....6S,2014ApJS..212....8S}. At both low and high redshifts ($z \sim 2 - 3$), the properties of {\OVI} associated with optically thick {\HI} absorbers concur with the physical properties of gas cooling in outflows in the CGM of star-forming galaxies \citep{2009ApJS..181..272G,2014ApJ...788..119L}. In these environments the {\OVI} can form at the narrow interface layers between cold clouds and a hotter phase of gas belonging to the halo, or the intragroup/intracluster medium \citep{1990MNRAS.244P..26B,2003ApJS..146..165S,2018MNRAS.475.3529N}. In the modelling of {\OVI} along both the sightlines, we find that collisional ionization in cooling plasma can explain the observed {\OVI} at solar metallicities. The {\OVI} and the BLA along both sightlines can be probing a warm ($T \sim 10^5$~K) transition temperature layer between cool gas of different column densities entrenched in the hot halo of galaxy G2 or G3 or hot intra-group gas. The chances of finding galaxies within impact parameter of $250$ kpc of {\OVI} system of equivalent width $\leq 0.2$~{\AA} in a group environment is high as compared to isolated galaxies \citep{2017ApJ...844...23P}. 

If instead, the {\OVI} along both sightlines are from the diffuse warm phase of the CGM of the nearest galaxy (which is G3), or constitutes warm intragroup gas, then the ionization fraction of {\OVI} given by the hybrid model, and the physical scales along and across the lines of sight can be used to determine the mass of this phase. Using equation 2 of \citet{2011Sci...334..948T} for estimating the gas mass for the mean {\OVI} column density, and CGM gas radius of $\log [\langle N(\OVI)\rangle /\cmsq] = 14.2$, and $R = 260$~kpc, \footnote{For a line of sight thickness of $L \sim 35$~kpc and a transverse separation of $l \sim 520$~kpc, the radial extent for a spherical distribution of gas is given by $R^2 = (L/2)^2~+~(l/2)^2$. See Fig. 3 of \citep{2014ApJ...784....5M} for a geometrical description of this expression.} respectively, and an {\OVI} covering fraction of $0.8$, we estimate a CGM or warm intragroup gas mass of $\gtrsim 4.5 \times 10^{9} \times$~Z$_\odot$/Z~M$_\odot$. The estimate is a conservative lower limit based on the $1\sigma$ lower limit of $T = 0.5 \times 10^5$~K in the hybrid models for the BLA - {\OVI} gas. The metallicity of the {\OVI} gas is poorly constrained due to the uncertainty in the temperature given by the BLA line widths. The mass estimate will be an order of magnitude higher, if the metallicity of the {\OVI} - BLA gas is one-tenth of solar, instead of solar. Similarly, at $T > 0.5 \times 10^5$~K, the observed {\OVI} will be recovered by the hybrid models at lower ionization parameters in which the ionization fraction of {\OVI} would be lower, thus raising the CGM/warm intragroup gas mass estimate.

\section{CONCLUSIONS}
In this work, we have presented the metallicity and ionization analysis, and information on the extended galaxy environment for an absorber at $z_{abs} = 0.399$ seen along two closely separated lines of sight towards the background quasars Q~$0107 - 025$A and Q~$0107 - 025$B . The key results are as follows:

\begin{enumerate}

    \item The twin lines of sight are probing a multiphase medium across transverse physical scales of $520$~kpc. Along sightline B, the absorber is a pLLS with $\log~[N(\HI)/\cmsq] = 16.82~{\pm}~0.06$, with concurrent detections of {\CII}, {\OII}, {\CIII}, {\OIII}, {\SiIII} and {\OVI}. Sightline A probes a lower column density environment with  $\log~[N(\HI)/\cmsq] = 14.23~{\pm}~0.05$, with the low ions, except {\CIII} and {\CIV} ($HST$/FOS), as non-detections. 
    
    \item The {\OVI} absorbers along both sightlines have comparable column densities of $\log (N/\cmsq) \sim 14.1 - 14.3$ and broad $b$-values of $\sim 33 - 52$~{\kms}. In comparison, the low ionization lines are narrower with $b \sim 11 -20$~{\kms} tracing cooler ($T \sim 10^4$~K) gas phases. 
    
    \item The sightline A shows two component {\HI}, {\CIII} and {\OVI} profiles spread over the same velocity. A photoionized phase with [C/H] = 0 and $n(\H) = 2.0 \times 10^{-4}$~{\cc} explains the {\HI}, {\CIII} and (also the {\CIV} from $HST$/FOS) in the central component. However, the {\OVI} in both components are consistent with collisionally ionized gas with $T \gtrsim 10^5$~K with the {\HI} in the offset component being a BLA with $b(\HI) = 68~\pm~18$~{\kms}. For the $T \sim 1.6 \times 10^5$~K given by the different $b$-values of the BLA and {\OVI} in this component, the observed column density of {\OVI} and {\CIII} are consistent with collisional ionization in a radiatively cooling plasma with [O/H] = [C/H] = $0$, similar to the metal abundance in the photoionized gas in this absorber. 
    
    \item In the pLLS observed along sightline B, the low and intermediate ions along with bulk of the {\HI} are tracing photoionized gas with $n_{\H} = 2.5 \times 10^{-3}$~{\cc} and [O/H] = [C/H] = $-1.1$. The {\OVI} from this phase is severely underproduced. A BLA with $\log~N[\HI/\cmsq] = 14.84~{\pm}~0.16$ and $b(\H) = 64~{\pm}~9$~{\kms} is required to explain the broad wings seen in the {\Lya} and {\Lyb} profile. The BLA and {\OVI} indicate the presence of gas with $T = 8.9_{-5.6}^{+5.6} \times 10^{4}$~K. The {\OVI} in the pLLS can be reproduced by photoionization in a more diffuse ($n_{\H} \sim 3.2 \times 10^{−5}$~{\cc}), higher ionization gas of similar metallicity ($1/10$-th solar) as the low ionization phase. Alternatively, the {\OVI} could be from a radiatively cooling collisionally ionized plasma with $T \lesssim 10^5$~K and [O/H] = $0$, similar to the metallicity in the {\OVI} gas along sightline A. 
    
    \item The sightlines are intercepting a galaxy group environment with 16 galaxies within a projected separation of $\rho < 5$~Mpc and velocity offset of $\Delta v < 500$~{\kms} from the mean absorber redshift. The sightlines are at projected separations of $\rho/R_{vir} \sim 1.1$ and $\rho/R_{vir} \sim 0.8$ from a galaxy (G3) of stellar mass $M_* \sim 10^{10.7}$~M$_\odot$, and luminosity $L_R \sim 0.07L_R^*$. The pLLS absorber along sightline B with an inferred metallicity of $1/10$-th solar happens close to projected major axis of this galaxy (azimuthal angle, $\Phi \sim 3^{\circ}$) consistent with the large {\HI} column a result of the line of sight piercing a cold accretion stream from the intragroup medium. The SFR $< 0.1$~M$_\odot$~yr$^{-1}$ of this galaxy is however inconsistent with the solar metallicity we derive for the low ionization gas along sightline A and for the {\OVI} along both sightlines. An alternate possibility is for the absorption along both sightlines to be tracing metals dispersed in a past bipolar outflow ($\sim 600$ Myr) from another galaxy (G2) at $\rho/R_{vir} \sim 1.2$ and $\rho/R_{vir} \sim 1.7$ with similar stellar mass and luminosity, but with a SFR $\sim 3$~M$_{\odot}$~yr$^{-1}$. 
    
    \item The {\OVI} along both sightlines are consistent with tracing radiatively cooling plasma in a collisionally ionized interface layer between a low ionization gas cloud and the hotter phase of the CGM of the galaxy G3, which is nearest in normalized impact parameter ($\rho/R_{vir}$), or the larger intragroup gas. Collisional ionization models of such a non-equilibrium scenario predict [O/H] $\sim 0$. Alternatives are for the {\OVI} to be tracing the warm diffuse CGM of the nearest galaxy (G3), or warm intragroup gas, in which case, the physical scales of $L \sim 35$~kpc and $l \sim 520$~kpc along and across the lines of sight of this gas phase yield an ionized gas mass lower limit of $\gtrsim 4.5 \times 10^{9} \times$~Z$_\odot$/Z~M$_\odot$. 
\end{enumerate}


\begin{landscape}
\begin{table}
    \centering
    \caption{Galaxy environment surrounding the absorbers at $z_{abs} = 0.399$ towards the twin sight lines A and B.}
    \label{tab:Galaxy}
    \begin{threeparttable}
    \resizebox{1.35\textwidth}{!}{%
    \begin{tabular}{lcccccccccccccccccr}
        \hline
        \multirow{4}{*}{Label} & \multirow{4}{*}{RA} & \multirow{4}{*}{DEC} & \multirow{4}{*}{m$_R$} & \multirow{4}{*}{z$_{gal}$} & \multicolumn{2}{c}{$\rho$ (Mpc)} & \multicolumn{2}{c}{$\Delta v$ (km/s)} & \multirow{4}{*}{M$_R$} & \multirow{4}{*}{log(L/L$^*$)} & \multirow{4}{*}{log(M$_{*}$/M$_\odot$)} & \multirow{4}{*}{R$_{vir}$} & \multirow{4}{*}{$\rho_A$/R$_{vir}$} & \multirow{4}{*}{$\rho_B$/R$_{vir}$} & \multirow{4}{*}{log(M$_{h}$/M$_\odot$)}  & \multirow{4}{*}{SF/SFR } & \multirow{3}{*}{SFR line}  & \multirow{4}{*}{Instrument}\\
        \cline{6-9}
         & & & & & A & B & A & B & & & & & & & & & & \\
         & (deg) & (deg) & & & & & & & & & & (kpc) & & & & (M$_\odot$/yr) & & \\ 
         (1) & (2) & (3) & (4) & (5) & (6) & (7) & (8) & (9) & (10) & (11) & (12) & (13) & (14) & (15) & (16) & (17) & (18) & (19)\\
        \hline
        G1 & 17.56048 & -2.32871 & 25.1 & 0.39954 & 0.121 & 0.318 & -12 & -83 & -16.5 & -2.6 & 8.7 & 87.9 & 1.3 & 3.5 & 11.0 & 0.1 & [O II] & MUSE\\
        G2 & 17.56194 & -2.32547 & 22.0 & 0.39935 & 0.179 & 0.249 & 28 & -42 & -19.6 & -1.4 & 10.0 & 143.2 & 1.2 & 1.7 & 11.6 & 2.9 & [O II] & GMOS\\
        G3 & 17.56718 & -2.32438 & 21.3 & 0.39882 & 0.276 & 0.200 & 142 & 71 & -20.3 & -1.1 & 10.6 & 257.2 & 1.0 & 0.7 & 12.4 & < 0.1 & H $\alpha$ & VIMOS\\
        G4 & 17.51073 & -2.34742 & 21.3 & 0.39992 & 0.915 & 1.288 & -92 & -163 & -20.3 & -1.1 & 9.9 & 136.3 & 6.6 & 9.3 & 11.6 & 1.8 & [O II] & GMOS\\
        G5 & 17.53755 & -2.38853 & 21.3 & 0.39882 & 1.164 & 1.566 & 143 & 73 & -20.3 & -1.1 & 10.4 & 193.2 & 5.9 & 8.0 & 12.0 & 3.0 & [O II] & GMOS\\
        G6 & 17.52869 & -2.38692 & 20.8 & 0.39930 & 1.197 & 1.611 & 40 & -31 & -20.8 & -0.9 & 10.6 & 232.2 & 5.0 & 6.8 & 12.3 & Yes & & DEIMOS\\
        G7 & 17.56568 & -2.27001 & 20.1 & 0.40054 & 1.214 & 0.861 & -226 & -296 & -21.5 & -0.6 & 11.0 & 428.5 & 2.8 & 1.9 & 13.1 & Yes & & DEIMOS\\
        G8 & 17.60223 & -2.37862 & 19.8 & 0.40053 & 1.305 & 1.425 & -223 & -294 & -21.9 & -0.4 & 11.2 & 834.3 & 1.5 & 1.6 & 13.9 & No & & VIMOS\\
        G9 & 17.58691 & -2.40884 & 19.7 & 0.39912 & 1.635 & 1.884 & 79 & 8 & -21.9 & -0.4 & 11.4 & 1357.3 & 1.1 & 1.3 & 14.6 & No & & GMOS\\
        G10 & 17.63162 & -2.36604 & 21.6 & 0.39922 & 1.643 & 1.604 & 57 & -13 & -20.0 & -1.2 & 10.7 & 279.9 & 5.8 & 5.6 & 12.5 & Yes & & DEIMOS\\
        G11 & 17.58439 & -2.41457 & 21.4 & 0.39993 & 1.722 & 1.985 & -96 & -167 & -20.2 & -1.1 & 9.9 & 138.1 & 12.3 & 14.2 & 11.6 & Yes & & VIMOS\\
        G12 & 17.48010 & -2.26885 & 23.9 & 0.39950 & 1.900 & 1.924 & -4 & -74 & -17.7 & -2.1 & 10.1 & 160.2 & 11.7 & 11.8 & 11.8 & Yes & & DEIMOS\\
        G13 & 17.51376 & -2.20793 & 22.3 & 0.39978 & 2.536 & 2.323 & -63 & -134 & -19.3 & -1.5 & 8.6 & 86.5 & 29.0 & 26.5 & 11.0 & Yes & & VIMOS\\
        G14 & 17.42589 & -2.35586 & 20.3 & 0.39945 & 2.559 & 2.884 & 7 & -64 & -21.3 & -0.7 & 11.1 & 602.1 & 4.2 & 4.7 & 13.5 & Yes & & VIMOS\\
        G15 & 17.62484 & -2.44852 & 22.1 & 0.40178 & 2.661 & 2.847 & -491 & -562 & -19.5 & -1.4 & 9.6 & 120.1 & 21.9 & 23.4 & 11.4 & 1.6 & H $\alpha$ & VIMOS\\
        G16 & 17.73452 & -2.20009 & 21.7 & 0.40180 & 4.339 & 3.939 & -496 & -567 & -20.0 & -1.2 & 9.2 & 102.5 & 41.8 & 38.0 & 11.2 & Yes & & VIMOS\\
 		\hline
    \end{tabular}
    }
    \begin{tablenotes}\footnotesize
    \item[(1)] Label of the 16 galaxies labelled as G1, G2 and so on. Information about G1 is from our analysis of the MUSE observations (Beckett {\etal}, in preparation) while rest are taken from \citet{2014yCat..74372017T} which are MOS observations.\\
    \item[(2)] RA of the galaxies in degrees.\\
    \item[(3)] DEC of the galaxies in degrees.\\
    \item[(4)] The apparent magnitude of the galaxies in R-band is taken from (Beckett {\etal}, in preparation) and \citet{2014yCat..74372017T}.\\
    \item[(5)] Redshift of the galaxies. \\
    \item[(6)] Impact parameters in kpc of galaxies with respect to the quasar line of sight A.\\
    \item[(7)] Impact parameters in kpc of galaxies with respect to the quasar line of sight B.\\
    \item[(8)] Velocity offset between the galaxies and the absorber at z$_{abs} = 0.39949$ towards line of sight A.\\
    \item[(9)] Velocity offset between the galaxies and the absorber at z$_{abs} = 0.39916$ towards line of sight B. \\
    \item[(10] By using cosmology of \citet{2014ApJ...794..135B}, we calculated absolute magnitude of the galaxies in the R-band. \\
    \item[(11)] The Schechter absolute magnitude of the L$^*$ galaxy \citep{1976ApJ...203..297S} in the R-band (M*$_R$) for the galaxies were calculated using relation in \citet{2005ApJ...631..126D}. This in turn gave the values of (L/L$^*$). \\
    \item[(12)] Stellar masses were estimated using the mass-to-light ratio tables as a function of colour, given in Table A7 of \citet{2003ApJS..149..289B}. These relationships have a scatter of 0.1-0.2 dex, with systematic shifts of up to 0.4 dex dependent on the assumed initial mass function.\\
    \item[(13)] The halo mass was converted into r$_{200}$ (kpc) using the halo mass by adopting the cosmology of \citet{2014ApJ...794..135B}.\\
    \item[(14)] The ratio of impact parameter of the galaxies with respect to sightline A and their virial radii. \\
    \item[(15)] The ratio of impact parameter of the galaxies with respect to sightline B and their virial radii. \\
    \item[(16)] Halo masses are estimated using the stellar-to-halo-mass relation given in \citet{2010ApJ...717..379B}. The uncertainty in halo mass is dominated by a $\sim$ 0.25 dex uncertainty in stellar mass, which corresponds to $\sim$ 25 $\%$ in R$_{vir}$. \\
    \item[(17) \& (18)] Star-formation rates (SFR) were estimated by converting H $\alpha$ or [O II] emission line luminosity to SFR with the formulae given in \citet{1998ARA&A..36..189K}, after applying 1 magnitude of extinction (approximately the mean for low-redshift galaxies as found by \citet{2002MNRAS.330..876C}).\\
    \item[(19)] This column mentions the spectrographs used for detecting galaxies and obtaining their spectra.
     \end{tablenotes}
    \end{threeparttable}
\end{table}
\end{landscape}

\newpage
\begin{landscape}
\begin{table}
    \centering
    \caption{Summary of ionization modelling results.}
    \label{tab:Summary_A&B}
    \begin{threeparttable}
    \begin{tabular}{lccccccccr}
    \hline
        Line of sight & $z_{abs}$ & Ionization phase & Model & log[N(\HI)/{\cmsq}] & T (K)  & $n(\H)$~({\cc}) & [X/H] & $N(\H)$~({\cmsq}) & $L$~(kpc) \\
        \hline
        \multirow{4}{*}{Q~$0107-025$A} & $0.39949$ & central low & PIE & $14.23 \pm 0.05$ & \textbf{$2.4_{-0.6}^{+0.6} \times 10^{4}$} & $2.0 \times 10^{-4}$ & $0.0$ & $4.09 \times 10^{17}$ & $0.7$  \\ 
         & & central high & ... & ... & $\leq 1.4 \times 10^{6}$ & ... & ... & ... & ... \\
         & & offset high & PIE + CIE & $13.52 \pm 0.05$ & $(0.5-1.5) \times 10^{5}$ & $\sim (0.4-0.7) \times 10^{-4}$ & $0.0$ & $3.25 \times 10^{18}$ & $19.5$  \\
         & & offset high & NECI & $13.52 \pm 0.05$ & $\sim 1.6 \times 10^5$ & ... & $0.0$ & ... & ...  \\
       \hline
       \multirow{9}{*}{Q~$0107-025$B} & $0.39916$ & central low & PIE & $16.82 \pm 0.06$ & $\sim 1.6 \times 10^{4}$ & $2.5 \times 10^{-3}$ & $-1.1$ & $1.66 \times 10^{19}$ & $2.2$  \\
        \cline{3-10}
        & & \multicolumn{8}{c}{Model I} \\
        \cline{3-10}
        & & central high & PIE & $14.84 \pm 0.16$ & $4 \times 10^{4}$ & $3.2 \times 10^{-5}$ & $-1.1$ & $2.18 \times 10^{19}$ & $221.0$  \\
        & & central high & NECI & $14.84 \pm 0.16$ & $8.9_{-5.6}^{+5.6} \times 10^{4}$ & ... & $0.0$ & ... & ...  \\
        & & central high & PIE + CIE & $14.84 \pm 0.16$ & $8.9_{-5.6}^{+5.6} \times 10^{4}$ & $2.0 \times 10^{-4}$ & $0.0$ & $4.14 \times 10^{19}$ & $67.3$  \\
        \cline{3-10}
        & & \multicolumn{8}{c}{Model II} \\
        \cline{3-10}
        & & central high & PIE & $14.90 \pm 0.23$ & $4 \times 10^{4}$ & $3.2 \times 10^{-5}$ & $-1.1$ & $2.51 \times 10^{19}$ & $253.7$  \\
        & & central high & NECI & $14.90 \pm 0.23$ & $4.3_{-3.3}^{+3.3} \times 10^{4}$ & ... & $0.0$ & ... & ...  \\
        & & central high & PIE + CIE & $14.90 \pm 0.23$ &  $7.8_{-6.8}^{+6.8} \times 10^{4}$ & $2.0 \times 10^{-4}$ & $0.0$ & $3.09 \times 10^{19}$ & $50.2$  \\
    \hline 
    \end{tabular}
     \begin{tablenotes}\footnotesize
    \textit{Note.} The columns show, from left to right: name of the quasar line of sight, the redshift of the absorber towards the corresponding line of sight, different phases probed by the absorber by single/multi-components, the equilibrium model adopted for ionization modelling, neutral hydrogen column density with $1\sigma$ error, temperature due to the different phases present, gas density of the absorber, metallicity of the absorber, total column density of the absorber and line of sight thickness of the absorber. The temperature listed is from the combined $b$-values of {\HI} and metal ions. For the high ionization phase in the central component towards sightline A, the associated {\HI} column density cannot be determined from the data. The upper limit on temperature is from the $b$-value of {\OVI}. We have adopted two models to explain the BLA component in absorption profiles of {\HI} where \textit{Model I} depicts BLA alone with pLLS component (2-component) and \textit{Model II} depicts BLA incorporated from both the 2-component and 3-component fit of {\HI}
     \end{tablenotes}
     \end{threeparttable}
\end{table}
\end{landscape}

\newpage
\section*{Acknowledgements}
Based on observations made with the NASA/ESA \textit{Hubble Space Telescope}, support for which was given by NASA through grant HST GO-14655 from the Space Telescope Science Institute. This work is based on observations collected at the European Southern Observatory under ESO programmes 086.A-0970, 087.A-0857 and 094.A-0131. This research has made use of the Keck Observatory Archive (KOA), which is operated by the W. M. Keck Observatory. This work is also based on observations obtained at the Gemini Observatory, which is operated by the Association of Universities for Research in Astronomy, Inc., under a cooperative agreement with the NSF on behalf of the Gemini partnership: the NSF (United States), the National Research Council (Canada), CONICYT (Chile), the Australian Research Council (Australia), Ministério da Ci\^encia, Tecnologia e Inovação (Brazil) and Ministerio de Ciencia, Tecnolog\'ia e Innovación Productiva (Argentina). We thank the people responsible for the tools and packages used in this work (Cloudy, VPFIT, Numpy).  

Support for this work was provided by SERB through grant number EMR/2017/002531 from the Department of Science \& Technology, Government of India. SM thanks the Alexander-von-Humboldt Foundation, Germany. AB acknowledges the support of a UK Science and Technology Facilities Council (STFC) PhD studentship through grant ST/S505365/1. SLM acknowledges the support of STFC grant ST/T000244/1.

\section{Data Availability}
The QSO spectroscopic data underlying this article can be accessed from the HST Spectroscopic Legacy archive \url{(https://archive.stsci.edu/hst/spectral_legacy/)} which is available in the public domain. The galaxy data is available in the public domain from \citet{2014yCat..74372017T} through the VizieR online catalogue \url{(https://vizier.u-strasbg.fr/)} with catalogue number: J/MNRAS/437/2017, and also in the ESO Science Archive Facility \url{(http://archive.eso.org/cms.html)}.



\bibliographystyle{mnras}
\bibliography{Biblio} 


\appendix
\section{Line of Sight towards LBQS~$0107-0232$}
\label{Appendix A}

\begin{figure}
\begin{center}
	\includegraphics[width=\columnwidth]{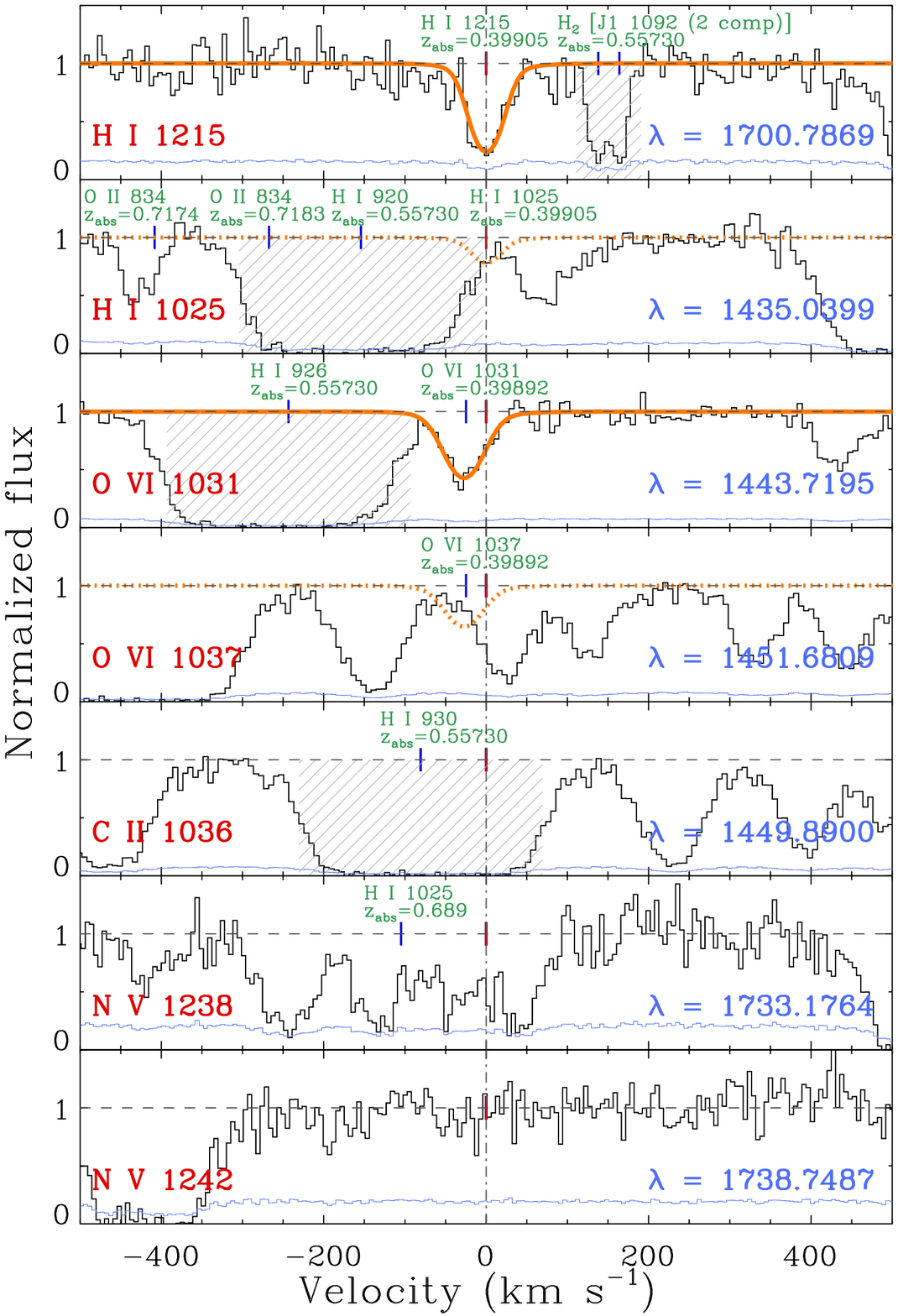}
    \caption{Figure similar to Fig.~\ref{fig:A_detections} covering {\HI} and some of the prominent metal lines at $z = 0.39905$ towards sightline LBQS~$0107-0232$. The identified contamination in each panel is labeled, and the contaminated regions are shaded. The \textit{solid orange} curves denote Voigt profile fits to {\Lya} and the feature close to the expected location of {\OVI}~$1031$. Based on these fits, the expected profiles of {\Lyb} and {\OVI}~$1037$ were synthesized and overlaid on the data (shown as \textit{dashed} curve. The feature at $-25$~{\kms} is inconsistent with being {\OVI}~$1031$~{\AA} at $z = 0.39905$.}
    \label{fig:third_sightline}
\end{center}
\end{figure}

The quasar LBQS~$0107-0232$ (sightline C) at z$_{em} = 0.726$ is at angular separations of $2.94^{\prime}$ and $1.94^{\prime}$ relative to sightlines A and B. These correspond to transverse separations of $1.2$~Mpc and $780$~kpc at $z = 0.399$ respectively, which is greater than the separation between A and B. We searched for absorption at $z \sim 0.399$ along sightline C. Due to lack of grating data, only a few lines are covered for this redshift. The systemplot is shown in Fig.~\ref{fig:third_sightline}. The locations where the key transitions are expected, including {\OVIdblt}, suffer contamination from lines associated with a sub-DLA at $z = 0.55730$ \citep{2013MNRAS.433..178C}. 

There is a $> 3\sigma$ feature at the expected location of {\Lya}. The region where the corresponding {\Lyb} is expected is contaminated by the {\HI}~$920$~{\AA} from the sub-DLA. Information on the higher order Lyman lines are not available as they fall blueward ($\lambda < 1420$~{\AA}) of the Lyman break from the sub-DLA. A single component free-fit to the {\Lya} yields $\log~[N/\cmsq] = 13.76~\pm~0.11$, $b = 22 \pm 7$~{\kms}, which is $0.5$~dex and $3$~dex lower compared to the {\HI} column density along lines of sight A and B respectively. We also identify two component molecular hydrogen line redward of {\Lya} associated with the high redshift sub-DLA. The absorption closest in velocity to the expected location of {\OVI}~$1031$~{\AA} is offset by $-25$~{\kms} from $v = 0$~{\kms} given by {\Lya}. We could not identify this feature with any of the metal lines or molecular hydrogen lines associated with the sub-DLA. Assuming it to be {\OVI}~$1031$ from $z = 0.399$, a free-fit yields $\log~[N/\cmsq] = 14.17~\pm~0.06$, $b = 27 \pm 4$~{\kms}, $v = -25~\pm~6$~{\kms}. Based on this the predicted {\OVI}~$1037$ profile is inconsistent with the data (see Fig.~\ref{fig:third_sightline}), suggesting that the absorption at $\lambda = 1442.3$~{\AA} may not be {\OVI}~$1031$. The {\NV}~$1242$~{\AA} falls at a clean region of the spectrum, where it is a non-detection. Since no conclusive measurements can be obtained for the metal absorption at $z = 0.399$, we have excluded this sightline from our main analysis.

\bsp	
\label{lastpage}
\end{document}